\documentclass[onecolumn,preprintnumbers,amsmath,
amssymb,nofootinbib,epsfig,bmamsfonts,yfonts,
superscriptaddress,notitlepage,11pt]{revtex4-1}
 \pdfoutput=1
\usepackage{graphicx}
\usepackage{amsmath,amssymb,amsthm,aas_macros}
\usepackage{latexsym,graphicx,color,subfigure,slashed,mathrsfs,verbatim,bbm,wasysym, pstricks,epsfig,url}
\usepackage{epstopdf}
\usepackage{nicefrac}
\usepackage{amssymb}
\usepackage{amsmath}
\usepackage{xcolor}

\DeclareMathOperator\arctanh{arctanh}

\newcommand{\beq}{\begin{eqnarray}}
\newcommand{\eeq}{\end{eqnarray}}

\newcommand{\grav}{grav}
\newcommand{\medium}{medium}
\newcommand{\dressed}{dressed}
\newcommand{\bh}{{bullet} }
\newcommand{\dd}{\text {d} }

\begin{document}

\title{{\Large Dynamical Friction in Interacting Relativistic Systems}}
\author{Andrey Katz}
\affiliation{TH Department, CERN, 1 Esplanade des Particules, 1211 Geneva 23, Switzerland}
\affiliation{Universit\'e de Gen\`eve, Department of Theoretical Physics and  Center for Astroparticle Physics (CAP), \\
	24 quai E. Ansermet, CH-1211, Geneva 4, Switzerland}
\author{Aleksi Kurkela}
\affiliation{TH Department, CERN, 1 Esplanade des Particules, 1211 Geneva 23, Switzerland}
\affiliation{Faculty of Science and Technology, University of Stavanger, 4036 Stavanger, Norway}
\author{Alexander Soloviev}
\affiliation{Institut f\"ur Theoretische Physik, Technische Universit\"at Wien,
	Wiedner Hauptstr. 8-10, A-1040 Vienna, Austria}
\date{\today}

\begin{abstract}
	\noindent
	We study dynamical friction in interacting relativistic systems with arbitrary
mean free paths and medium constituent masses. Our novel framework recovers the
known limits of ideal gas and ideal fluid when the mean free path goes to
infinity or zero, respectively, and allows for a smooth interpolation between
these limits. We find that in an infinite system the drag force can be
expressed as a sum of ideal-gas-like and ideal-fluid-like contributions
leading to a finite friction even at subsonic velocities. This simple picture
receives corrections in any finite system and the corrections become especially
significant for a projectile moving at a velocity $v$ close to the speed of
sound $v\approx c_s$. These corrections smoothen the ideal
fluid discontinuity around the speed of sound and render the drag force a continuous function
of velocity. We show that these corrections can be computed to a good
approximation within effective theory of viscous fluid dynamics.
\end{abstract}

\preprint{CERN-TH-2019-088}

\maketitle

\section{Introduction}
\label{sec:intro}


Dynamical friction---or how a projectile that moves through a medium is slowed
down by the gravitational interaction with the medium---is a classic problem
that has many applications, primarily in astrophysics. It was first addressed by
Chandrasekhar~\cite{1943ApJ....97..255C}, who considered the gravitational force
exerted on a projectile that moves non-relativistically through a gas of heavy,
non-interacting particles. The primary motivation of Chandrasekhar's work was to
understand the drag force exerted on a star in a ``gas'' of other stars. 
However, there are many modern questions in astro-particle and
beyond-the-standard-model physics involving the gravitational interaction of a
projectile with its surrounding medium---such as, \emph{e.g.}, the propagation
of primordial black holes or other dark-matter candidates through neutron
stars~\cite{Capela:2012jz,Capela:2013yf,Pani:2014rca,Capela:2014qea}---which
cannot be reduced to non-relativistic movement in a dilute gas, motivating
further analysis of dynamical friction in more general systems.

Chandrasekhar's result has indeed been generalized to many different systems.
For example, the first analysis of the relativistic projectile propagating
through a non-interacting gas of photons was performed
in~\cite{1994MNRAS.270..205S}. However, the interpolation between these two
limits of non-relativistic and ultra-relativistic media has so far remained
unknown. Besides mere kinematics, dynamical friction depends crucially also on
the self-interaction of the medium constituents. If the interaction between the
medium constituents is strong enough, it is more appropriate to describe the
medium as an ideal fluid rather than as an ideal gas---a case studied
in~\cite{1980ApJ...240...20R,1971ApJ...165....1R,1964SvA.....8...23D,Ostriker:1998fa,2007MNRAS.382..826B}. Again, how one interpolates between the two limits of ideal gas and ideal fluid in a system with finite mean free path remains an open question.

In this work we develop a simple diagrammatic field-theory formalism to
evaluate dynamical friction in a wide class of relativistic interacting systems
with a finite mean free path, $l_{\rm mfp}$. Dynamical friction arises from the
projectile moving up the gravitational potential it itself created by
gravitationally perturbing the medium through which it propagates. The shape of
the wake---and hence its gravitational field---depends on the material
properties of the medium that in an interacting system depend on the length
scale considered. At short length scales $\Delta x \ll l_{\rm mfp}$, the medium
appears as an ideal gas and corrections may be computed in powers of $\Delta
x/l_{\rm mfp}$. In the opposite limit of long length scales $\Delta x \gg l_{\rm
	mfp}$, the interacting medium behaves as an ideal fluid, a picture that can be
systematically improved by taking into account higher-order "viscous"
corrections appearing in powers of $l_{\rm mfp}/\Delta x$ in fluid-dynamic
gradient expansion~\cite{Baier:2007ix}.

 Due to the long-distance nature of the gravitational force, all length scales
contribute equally to dynamical friction. In an interacting medium where the
mean free path fits inside the medium $R_{\rm max} > l_{\rm mfp}$ but is larger
than the size of the projectile $R_{\rm min} < l_{\rm mfp}$, the drag force
receives both ideal-gas- and ideal-fluid-like contributions 
\begin{equation}
 F = c_{\rm gas} \log\left( \frac{R_{\rm min}}{l_{\rm mfp}}\right)+ c_X + c_{\rm
	fluid} \log\left( \frac{l_{\rm mfp}}{R_{\rm max}}\right), 
\end{equation} 
where the logarithmically enhanced terms arise from all scales much shorter
($c_{\rm gas}$)  or much longer ($c_{\rm fluid}$) than the mean free path. These
logarithmically enhanced terms are universal in the sense that they depend only
on few characteristics of the medium: the ideal-fluid-like contribution depends
only on the speed of sound, whereas the ideal-gas-like contribution depends on
an integral moment of the velocity distribution. All the information about the
interactions can be encapsulated in a single subleading-log contribution $c_X$.

We compute the drag force using a field-theory approach in which
the force---to leading order in Newton's constant---is given 
by the in-medium-dressed graviton propagator (see Fig.~\ref{fig1}). The computation therefore boils down 
to finding the graviton polarization tensor that dresses the tree-level graviton propagator. The polarization tensor that is given by the Green function of the energy-momentum tensor characterizes the medium and is known in variety of media. Here, we consider freely streaming gas, viscous fluid dynamics, and interacting kinetic theory in relaxation-time 
 approximation~\cite{Baier:2007ix,Romatschke:2015gic}.

		In the limit of ideal fluid dynamics, the drag force vanishes exactly for
		subsonic projectiles and the force has a discontinuity when crossing the sound barrier $v = c_s$ \cite{1980ApJ...240...20R}. 
		We find that in the fluid-dynamic limit, dynamical friction can be
understood as emission of real ``on-shell" phonons forming a Mach cone
transporting energy far away from the projectile.  This explains in an intuitive
way the lack of dynamical friction for a subsonic projectile that has
phase-space to excite only ``virtual" phonons which do not propagate far. We
further find that dissipative, viscous effects are always large for projectile
velocities near the speed of sound $v\approx c_s$. For these velocities a
description of the medium in terms of ideal fluid is never applicable and
viscous effects always lead to $\mathcal{O}(1)$ corrections. These corrections
smoothen the discontinuity of the drag force in any system with a finite
interaction rate (for other effects removing the discontinuity, see \cite{Ostriker:1998fa,Berezhiani:2019pzd}). By computing these subleading-log contributions in interacting
kinetic theory and in viscous fluid dynamics we find that they can be well
approximated by the latter.

We note that also our treatment makes two assumptions to simplify the problem.
First, we will restrict ourselves to a regime, where the accretion onto the
projectile that propagates through the medium is negligible. This is always the
case if the radius of the bullet is much bigger than the accretion radius,
however, this condition can also be satisfied in a broader framework. Clearly
this is not always the case,\emph{e.g.}, if the projectile is a non-relativistic
black hole. However, it has been previously shown in explicit numerical
simulations that even if the accretion effects are maximized, they never dominate over 
dynamical friction~\cite{1989ApJ...336..313P}. 
Second, we
restrict ourselves to the regime of linearized gravity. We will
briefly explain, however, how our formalism can potentially be used to capture the
post-Newtonian corrections.\footnote{For earlier works that address the
	post-Newtonian corrections to the dynamical friction
	see~\cite{1969ApJ...155..687L,1989ApJ...336..313P}.}

Our paper is structured as follows. In Sec~\ref{sec:formalism} we present our
formalism in detail and explain the logic behind the calculation. The
calculation itself is performed in Sec~\ref{sec:calculation}. We begin by
analyzing the fluid-dynamical case, starting from  the well-known inviscid fluid and
then continuing to the more complicated scenario with non-vanishing
viscosity. We then move on to the ideal-gas regime, reproducing the
well-known results of the non-relativistic and ultra-relativistic approximations.  Finally we perform the full calculation
for the interacting gas in relaxation time approximation that interpolates between all of these cases. The boundary
effects are very briefly addressed in Sec~\ref{sec:boundary}. In the last
section we draw our conclusions and discuss the application of our results for
real-words problems.


\section{Description of the Formalism}
\label{sec:formalism}


Our main objective is to calculate the drag force, exerted on gravitating point
source---or the \emph{bullet} from here onwards---that propagates through
yet-to-be-specified static and infinite medium. In the linearized regime, the
four-force $F^{\mu}$ exerted on the bullet of mass $m_b$ by the gravitational
field of its own wake $h^{wake}$ can be written as
\begin{align}\label{eq:formalforce}
F^\mu = \frac{\text{d}p^\mu}{\text{d}t}= -\frac{m}{\gamma}
\Gamma^\mu_{\phantom{\mu}\alpha\beta} (h^{wake})u^\alpha u^\beta,
\end{align}
where $u^\mu = \gamma (1, \vec v)$ is the four-velocity of the bullet. 
The zero component corresponds to the rate of energy loss $F^0 = \frac{\dd
	E}{\dd t}$. If the projectile in question has dimensions comparable to its
Schwarzschild radius, this approximation breaks down at short distances where
non-linear  effects in gravity start to play a non-negligible role. However,
this is expected to happen at distances strictly smaller than the accretion
radius \cite{1989ApJ...336..313P} where our calculation is insufficient anyways.
Therefore we shall adopt this approximation through out the paper.

\subsection{The wake}
We start by deriving a formal expression of the gravitational field of the
bullet's wake in a form that makes minimal assumptions about the structure of
the medium. We assume that before the medium is perturbed by the projectile, the
medium is homogeneous and isotropic such that its unperturbed energy-momentum
tensor reads\footnote{Our metric in this paper is ``mostly plus'', namely $(-+++)$. } 
\beq\label{eq:unperturbedtmunu}
T_{\mu \nu}^{medium} = \left(
\begin{array}{cccc}
	\rho & 0 & 0 & 0 \\
	0 & P & 0 & 0 \\
	0 & 0 & P & 0 \\
	0 & 0 & 0 & P
\end{array}
\right)
\eeq  
with energy density $\rho$ and pressure $P$. 
We assume that the 
unperturbed background metric can be treated as locally flat $g^{(0)}_{\mu\nu}= \eta_{\mu\nu}$.
Note however that for strongly self-gravitating objects this locally flat frame 
can be significantly different from the observer's frame.
 
 For large separations, the projectile affects the medium by inducing a linear perturbation of
the gravitational field
\begin{equation}
g_{\mu\nu} = 
\eta_{\mu\nu} + h^\bh_{\mu \nu}.
\end{equation}
The gravitational perturbation $h^\bh_{\mu\nu}$ caused by the bullet is
conveniently found by convoluting the
energy-momentum tensor of the bullet with the Green function of the
linearized Einstein equation, \emph{i.e.}, with the graviton propagator%
\footnote{Here, we work in the harmonic (de Donder) gauge, 
see appendix~\ref{App:graviton} for details. }
\begin{equation}
h^{\bh}_{\mu \nu}(x) = \int d^4 x' G^{\grav}_{\mu \nu, \alpha\beta}(x,x') T^{\alpha\beta}_{\bh}(x').
\end{equation}
Assuming that the bullet of mass $m_b$ moves with a constant velocity $v$
through the medium  in $z$-direction, the energy momentum tensor of the bullet
and its Fourier transform  can be written as\footnote{We work with the
	convention $f(x) = \int \frac{d^4 k}{(2\pi)^4} e^{i k\cdot x} f(k)$. Further, we
	will be interested in retarded Green functions, and therefore to pick the
	retarded ordering, we must take $\omega \equiv k^0$ to have a small positive
	imaginary part $\omega+i\epsilon$. }
\begin{align}
T_{\bh}^{\mu \nu}(t, \vec x) &= \frac{m_b}{\gamma} u^\mu u^\nu \delta(z-vt) \delta^2(x_\perp),  \\ 
\tilde T^{\mu \nu}_\bh(\omega, \vec k) &= \frac{2 \pi m_b}{\gamma} u^\mu u^\nu \delta (\omega - k^z v),\label{eq:tmunubullet}
\end{align}
 where $u^\mu = \gamma(1, 0, 0, v)$ is the four-velocity vector of the bullet.


The linear perturbation in the gravitational field caused by the bullet will in
turn cause a linear perturbation in the background medium. How this perturbation
evolves in time depends on the material properties of the medium. For now, we
will not specify the properties of the medium. However, in general, the medium
response of the energy-momentum tensor $\delta T^{\mu\nu}$, 
can be  characterized by its retarded Green function $G_{\medium}$
\begin{align}
\label{eq:twake}
\delta T_{wake}^{\mu \nu}(x) =  \int d^4 x' G_{\medium}^{\mu\nu,\alpha\beta}(x',x)h^{\bh}_{\alpha\beta}(x').
\end{align}
The response function depends on the microscopic dynamical properties of the
medium and Eq.~\eqref{eq:twake} needs to be eventually supplemented with an
appropriate Green function characterizing the medium. The Green function is
known is some special cases, \emph{e.g.}, in free-streaming gas and in
ideal liquid (or even in some quantum field theories~\cite{Hartnoll:2005ju}); 
we reproduce these
limits in the appendices and extend the calculation of the Green function to
massive interactive kinetic theory.

Finally, the wake itself will have its own gravitational field, which can again
be computed using the Green function of the linearized Einstein equation
\begin{align}
h^{wake}_{\mu \nu}(x) = \int d^4 x' G^{\grav}_{\mu \nu, \alpha\beta}(x,x') T^{\alpha\beta}_{wake}(x').
\end{align}
That is, $h^{wake}$ represents the gravitational field that was created by a
wake that was created by a gravitational field that was created by the bullet. 
In Fourier space, where the convolution becomes a simple product of the Green
functions, the gravitational field of the wake has a particularly simple form
\begin{align}
h^{wake}_{\mu \nu}(\omega, \vec k) =  G^{\dressed}_{\mu \nu, \alpha\beta}(\omega,\vec k) T^{\alpha \beta}_{\bh}(\omega, \vec k)
\label{Eq:hdress}
\end{align}
in terms of the retarded in-medium dressed graviton propagator
\begin{align}\label{eq:Gdressed}
G^{\dressed}_{\mu \nu, \alpha\beta}(\omega,\vec  k) = G^{\grav}_{\mu \nu, \gamma \delta} (\omega,\vec  k)G_{\medium}^{\gamma  \delta, \rho \sigma}(\omega,\vec  k)G^{\grav}_{\rho \sigma, \alpha\beta}(\omega,\vec  k).
\end{align}

\subsection{The force}
Once we know the expression for the gravitational field of the wake, the force that is exerted on the
bullet becomes straightforward to calculate.
Manifestly, Eq.~\eqref{eq:formalforce}  reads in components
\begin{align}\label{eq:forceChrist}
F^z&=-m\gamma(\Gamma^z_{\phantom{\mu}00}+2v \Gamma^z_{\phantom{\mu}0z}+v^2\Gamma^z_{\phantom{\mu}zz})
=-m\gamma\left(-\frac{1}{2}\partial_z h^{wake}_{00}+\partial_0 h^{wake}_{0z}+(v\partial_0 +v^2\frac{1}{2}\partial_z) h^{wake}_{zz}\right), 
\end{align}
where the above expression is evaluated along the worldline of the bullet
$z=vt$. The expression has a particularly simple interpretation in the
non-relativistic limit, where only the first term contributes and $h_{00}$
can be identified as the Newtonian gravitational potential: if the
gravitational potential has a gradient along the direction of the movement of
the bullet, the bullet does work climbing up the potential leading to energy
loss.

The gradients of the different components of the gravitational-field
perturbation are computed by inserting the energy-momentum
tensor~\eqref{eq:tmunubullet} to Eq.~\eqref{Eq:hdress} and performing the
inverse Fourier transform
\begin{align}\label{eq:wakeder}
\partial_\rho h^{wake}_{\mu \nu}(x) &
= i  \frac{m_b u^\alpha u^\beta}{\gamma} \int \frac{d^4 k}{(2\pi)^4} e^{i k\cdot x} 
k_\rho  G^{dressed}_{\mu \nu,\alpha \beta}(k)2\pi \delta(\omega-k_z v),
\end{align}
which has a simple form along the worldline of the bullet
\begin{align}
\label{eq:wakeder2}
\partial_\rho h^{wake}_{\mu \nu}(z = vt) &=  i  \frac{m_b u^\alpha u^\beta}{\gamma} \int \frac{\text{d}^2 k_\perp}{(2\pi)^2} \frac{\text{d}k_z}{2\pi}  k_\rho G^{dressed}_{\mu\nu,\alpha \beta}(\omega = v k_z
+i\epsilon
, \vec k)~.
\end{align} 
The $+i\epsilon$ prescription chooses the appropriate boundary condition 
for the retarded propagator. 

\begin{figure}
	\centering
	\includegraphics[width = 0.5\textwidth]{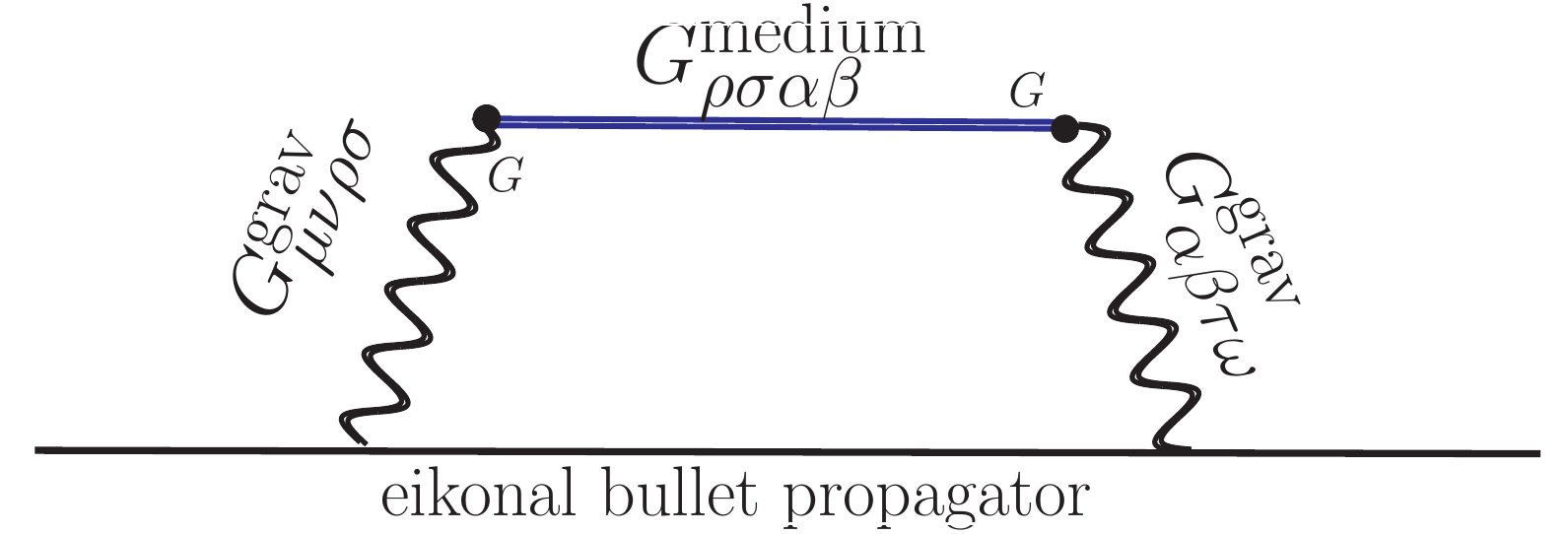}
	\caption{Diagrammatic presentation of the calculation of the drag force of
		\eqref{eq:fzintegral}. It should be understood as a mixing
		of the linearized graviton with the excitation of medium. In the limiting cases
		of the hydrodynamics and the free streaming processes, these excitations
		becomes the sound and shear modes or the particle excitations, respectively. }
	\label{fig1}
\end{figure}

From Eq.~\eqref{eq:forceChrist} it is clear that the only relevant cases for our
calculation are either $\rho = 0$ or $\rho = z$. In both of these cases the
integral over $dk_z$ in Eq.~\eqref{eq:wakeder2} can be further simplified by
dividing the right-hand side into two equal parts and changing the integration
variable $k_z \rightarrow - k_z$ in the latter part. As long as the background
over which $G^{\dressed}$ is computed is (time) translation invariant---as it is
in all the cases we are considering---changing the sign of the real part of  the
frequency turns a retarded Green function into an advanced one, which is given by the
complex conjugate of the retarded function. In all the cases that we are interested in, the retarded function
is multiplied by a single power of $k_z$, such that we get
\begin{align}
\int \frac{\text{d}k_z}{2\pi} i k_z G^{dressed}_{\mu\nu,\alpha \beta}&= \int \frac{\text{d}k_z}{2\pi}  \frac{i k_z}{2} (G^{dressed}_{\mu\nu,\alpha \beta}-(G^{dressed}_{\mu\nu,\alpha \beta})^*)
 = - {\rm Im} \int\frac{\text{d}k_z}{2\pi} k_z  G^{dressed}_{\mu\nu,\alpha \beta}.
\end{align} 

With a little bit of algebra one can show that the force is then given simply by
\begin{align}  \label{eq:fzintegral}
F^z=-\frac{m^2_{\rm bullet}}{2 \gamma^2} \text{Im} \left[ \int \frac{\text{d}^2
	k_\perp}{(2\pi)^2} \frac{\text{d}k_z}{2\pi}k_z  G^{\dressed} \right],
\end{align}
with
\begin{equation}
\label{contracts}
G^{\dressed} \equiv  \left[ G^{dressed}_{\mu\nu,\alpha \beta}\right] u^\mu u^\nu u^\alpha u^\beta,
\end{equation}
where we additionally used that $(-v\omega +v^2\frac{1}{2}k_z) = -\frac{v^2}{2}k_z$ and the explicit structure 
of the vector $u^\mu$ as it appears below Eq~\eqref{eq:tmunubullet}.

This expression has a simple diagrammatical representation, shown in
Fig.~\ref{fig1}. The dressed graviton propagator of (\ref{eq:Gdressed}) is
connected to the \emph{eikonal} propagator of the bullet (\ref{eq:tmunubullet})
which forces the frequency $\omega = v k_z$.


\subsection{Scale Separation and the Structure of the Drag Force}
In the previous section we developed the machinery to calculate the drag force
when the Green function of the energy-momentum tensor in medium is known. We now move on
to discuss the generic properties of the Green functions and their
implications to the structure of the drag force.

Irrespective of the microscopic details, any interacting matter with a finite
mean free path $l_{\rm mfp}$ admits an effective late-time ($\Delta t \gg \tau_{\rm scat}$), long-distance ($\Delta x \gg
l_{\rm mfp}$) description of the energy-momentum tensor in terms of relativistic
fluid dynamics~\cite{Baier:2007ix}. Relativistic fluid dynamics is based on a
gradient expansion of the energy-momentum tensor around its equilibrium value
\begin{align}
T_{\mu \nu} &= T^{\rm eq}_{\mu\nu}+ \Pi_{\mu \nu}, \\
T^{\rm eq}_{\mu \nu} &= (\rho + P) u^{\rm fluid}_\mu u^{\rm fluid}_\nu + P g_{\mu \nu},
\end{align}
where $u^{\rm fluid}_\mu$ is the rest frame of the fluid and the viscous shear-stress tensor,
$\Pi_{\mu \nu}$, contains all possible terms consistent with symmetries
constructed from the macroscopic quantities $P$, $\rho$, and $u_{\rm fluid}$ up
to a given order in gradients. For ideal fluid dynamics the shear-stress tensor vanishes $\Pi_{\mu \nu} = 0$,
whereas in first order viscous fluid dynamics the viscous shear-stress tensor is given by
\begin{align}
\label{hydro}
\Pi_{\mu \nu} &= -2\eta \sigma_{\mu\nu}-\zeta\nabla_\alpha u_{\rm fluid}^\alpha\Delta_{\mu\nu},\nonumber\\
\sigma^{\alpha\beta}&=\frac{1}{2}\Delta^{\mu\alpha}\Delta^{\nu\beta}\Big{(}\nabla_\mu u^{\rm fluid}_\nu+\nabla_\nu u^{\rm fluid}_\mu -\frac{2}{3}\nabla_\alpha u_{\rm fluid}^\alpha g_{\mu\nu}\Big{)}.
\end{align}
where the shear $\eta$ and bulk viscosities $\zeta$ are low-energy constants that
describe the dissipative properties of the medium, and the projector on the subspace orthogonal to the flow velocity is given by $\Delta^{\mu\nu}\equiv u^\mu u^\nu+g^{\mu\nu}$. The higher-order corrections
to ideal fluid dynamics are suppressed by powers of $l_{\rm mfp}/\Delta x \sim k
l_{\rm mfp} $. 

A simple dimensional analysis shows that the force in the ideal-fluid regime must be logarithmically divergent. The force in~\eqref{eq:fzintegral} must be parametrically 
\begin{align}
\label{eq:param}
F^z \sim m_{\rm bullet}^2 G^2 \int \frac{d^4 k}{k^4} \times (\rho + P),
\end{align} 
where the $G^2/k^4$ arises from the graviton propagator and is a reflection of the
long-distance nature of the gravitational Coulomb potential. In absence of other
available scales, the ideal fluid-dynamic energy-momentum tensor must be
$\mathcal{O}(\rho + P)$ (see Appendix \ref{App:fluid} for explicit expressions).
As is evident from~\eqref{eq:param}, all distance scales, $1/k$, contribute
equally to the force. This leads to a logarithmic divergence when integral over all scales $dk$ it taken. At large distance
scales (corresponding to small wave numbers $k$), the divergence is regulated by
dimensions of the medium $R_{\rm max}$. At small distance scales (large $k$),
the divergences is regulated either by size of the bullet or by its accretion
radius (whichever is bigger), $R_{\rm min}$, if the bullet is larger in size than
the mean free path $l_{\rm mfp}$.
 
The higher-order corrections to the ideal fluid dynamics arise from terms with
increasing number of gradients leading to corrections that become larger as
$k$ increases
\begin{align}
\label{param2}
F^z \sim m_{\rm bullet}^2 G^2 \int \frac{d^4 k}{k^4} \times (\rho + P) \left( 1 + \mathcal{O}(l_{\rm mfp} k) + \mathcal{O}\left( (l_{\rm mfp} k)^2 \right)+ \ldots \right).
\end{align} 
The viscous corrections have stronger short-distance divergences (but are
convergent at large distances), and they become increasingly more sensitive to
the short-distance cutoff. If the short-distance cut-off is larger than the mean
free path $R_{\rm min} > l_{\rm mfp}$, corrections to ideal fluid dynamics can
be computed in hydrodynamical approximation to desired accuracy: in Sec
\ref{sec:viscous_fluid} we provide the first viscous correction. However,
if the intrinsic scale of the bullet is smaller than the mean free path $R_{\rm
	min} < l_{\rm mfp}$, the gradient expansion breaks down at scale $k \sim
1/l_{\rm mfp}$.

At scales shorter than the mean free path $k > 1/l_{\rm mfp}$ the matter behaves
approximately as free-streaming. Also in the ideal gas, the drag force acquires
equal contributions from all scales leading to logarithmic divergence. Now the
interaction corrections come with increasing powers of $1/kl_{\rm mfp}$ (see
\emph{e.g.}~\cite{Kurkela:2019kip}). Now the higher order terms are convergent at short
distances but have increasingly strong long-distance divergences. That is, if
the extent of the system is smaller than the mean free path $R_{\rm min} <
l_{\rm mfp}$, the force can be computed in perturbative series of scattering
corrections.

In systems where the mean free path fits inside the dimensions of the system the
force gets contributions from all scales, the energy-momentum tensor can be
approximated by fluid dynamics in the large-wave-length region and by free-streaming in the short-density region, which both lead to logarithmic
divergences which are regulated by the dimension of the system and by the
breakdown scale of the respective approximations
\beq\label{eq:fullparametrization}
F^z =  c_{\rm gas} \log\left( \frac{R_{\rm min}}{l_{\rm mfp}}\right)+ c_X + c_{\rm
	fluid} \log\left( \frac{l_{\rm mfp}}{R_{\rm max}}\right)~.
\eeq 

The logarithmically enhanced coefficients $c_{\rm fluid}$ and $c_{\rm gas}$ can be computed in ideal
fluid dynamics and in free streaming systems, respectively. The contribution 
arising from scale $k\sim 1/l_{\rm mfp}$ gives rise to a subleading-log contribution, $c_X$, with no small
expansion parameters.

This cross-over regime needs to be calculated in the full
interacting kinetic theory, and it is sensitive to the microscopic details of
the interactions. 
In a system with large scale separation between the geometric
dimensions of the system and the mean free path, the ideal fluid and
free-streaming contributions are logarithmically enhanced compared to the
cross-over region arising from a single scale. However, as we will see in Section \ref{sec:breakdown}, this picture breaks down for $v\approx c_s$, irrespective of the value of $l_{\rm mfp}$.

  
\section{Calculation of the Drag Force} 
\label{sec:calculation}

We now apply the formalism that we developed in the previous section to compute
the dynamical friction in different regimes. We first start with a calculation
in the fluid-dynamic regime. While the ideal-fluid case is extensively covered in
literature, we provide the first results for the viscous fluid. We then discuss
the free streaming limit and extend the existing results by going beyond
non-relativistic and ultra-relativistic approximations. Finally, we compute the
drag force in a simple kinetic theory model that behaves as a free-streaming gas
at short distance scales but exhibits fluid-dynamical behavior at large distance
scales.

\subsection{Fluid-dynamic limit} 
The form of retarded
fluid-dynamic energy-momentum correlation function is well known;  we briefly
show this derivation in Appendix~\ref{App:fluid}. Using these results we can
straightforwardly write the dressed (scalar mode) graviton propagator 
\begin{align}
G^{dressed}(\omega= v k^z) &= \left( \frac{8\pi G  \gamma^2}{-(\omega + i \epsilon)^2 + 
	\vec
	k^2}\right)^2\frac{f^{\rm ideal}_{\rm sound}(\omega, \vec
	k)}{-(\omega+i\epsilon)^2 + c_s^2 k^2}\Bigg|_{\omega= vk^z} 
\\&= \left(
\frac{8 \pi G \gamma^2}{(1- v^2 )k_z^2 + \vec k_\perp^2 }\right)^2 \frac{f_{\rm
		sound}^{\rm ideal} (\vec k),}{(c_s^2 - v^2 )k_z^2 + c_s^2 \vec k_\perp^2 - i
	\epsilon v k_z} . \label{Eq:Gdress}
\end{align}
The squared expression arises from the two graviton propagators. In ideal fluid
dynamics, the perturbations of the energy-momentum tensor propagate by forming
long-lived longitudinal density perturbations, \emph{i.e.}, sound. The second
propagator is a phonon---or sound-mode---propagator, characterized by the speed of
sound $c_s^2 = \textstyle{dP/d\rho}$.  The numerator $f^{\rm ideal}_{\rm sound}$
reflects the coupling between gravity and sound modes and  comes from the
contractions of~\eqref{eq:Gdressed} and~\eqref{contracts} with the sound-mode
propagator of (\ref{Ghydro})
\beq
f_{\rm sound}^{\rm ideal} =  \frac{\rho +P}{2} \Bigg[ \frac{16 k_z^2 v^2 \omega^2}{k^2} + 
\Big(\omega (8 k_z v + (v^2 - 3)\omega) - k^2 (v^2 + 1) \Big( c_s^2 (v^2 -3) -v^2 -1
\Big)
\Big)\Bigg]
\eeq
As per~\eqref{eq:fzintegral}, the drag force arises from the imaginary part of
the above expression. As both the graviton and sound modes are long-lived, the
above expression is real up to the $i \epsilon$ description. Then, an imaginary
contribution can arise only when the the propagators go on shell, corresponding
to a production of a long-lived mode. Because $(1-v^2)>0$, the graviton propagators will never go on shell and the propagator is an
analytic function of $k^z$ for all $v$ along the real $k^z$-axis. The same is true for the
sound mode as long as $v < c_s$, and for subsonic velocities the propagator is
perfectly analytic and the bullet feels no drag force in the inviscid fluid, in
full agreement with~\cite{1980ApJ...240...20R}. That is, the gravitational field
of the bullet excites only off-shell excitations in the fluid---virtual
phonons--- that cannot propagate far and make up the wake trailing the bullet.
There is no phase space available to excite a real on-shell phonons that could
carry energy asymptotically far away from the bullet, and as energy cannot be
given to long lived modes, there can be no total energy loss on the bullet
because of energy conservation.

However, when the bullet moves with a supersonic velocity, namely $v^2-c_s^2 > 0$, 
the phase space opens up for the sound mode propagator to go on shell when 
\begin{align}
k^z = k^z_{sound} =  \pm \frac{c_s k_\perp}{\sqrt{v^2 - c_s^2}},
\label{Eq:pole}
\end{align}
leading to an imaginary part arising from the $i \epsilon$ prescription
\begin{align}
\frac{1}{(c_s^2 - v^2 )k_z^2 + c_s^2 \vec k_\perp^2 - i \epsilon v k_z} = 
{\cal P} \left( \frac{1}{(c_s^2 - v^2 )k_z^2 + c_s^2 \vec k_\perp^2 ) }\right) + 
i \pi \delta\left((c_s^2 - v^2 )k_z^2 + c_s^2 \vec k_\perp^2\right),
\label{Eq:principal}
\end{align}
where ${\cal P}$ denotes the real principal value integral.
Inserting~\eqref{Eq:principal} to~\eqref{Eq:Gdress} and further to~\eqref{eq:fzintegral}
we get a simple form for the drag force
\begin{align}
\label{ideal_fluid}
\frac{d F^z}{d \log k_\perp}= 
-\theta(v-c_s)m_b^2 4 \pi G\gamma^2 \frac{ (1+v^2)^2 (P+\rho)}{v^2}
\end{align}
in agreement with~\cite{2007MNRAS.382..826B}.

That is, we see that the energy loss arises from transferring energy from the
bullet to the long-lived asymptotic states---which in the case of fluid dynamics
are the sound modes---that can transport the energy far away from the bullet.
The condition~\eqref{Eq:pole} equivalently reads
\begin{equation}
\label{openingangle}
\cos\theta = k_z/|k| = c_s/v
\end{equation}
which is the angle in which the phonons are created and the unidirectional
production of the sound waves is nothing but the sonic boom created by the
bullet with the opening angle $\theta$. 

\subsubsection{Viscous fluid}
\label{sec:viscous_fluid}

We are now ready to analyze a more realistic situation of the drag force in a viscous fluid. As we will immediately 
see, the drag force in a viscous fluid does not vanish in the subsonic regime.

In a viscous fluid, the gravitational field excites both sound (\ref{Ghydro}) and shear modes (\ref{shearmode})  (but not the tensor mode (\ref{tensormode})) and using the fluid-dynamical Green function from Appendix \ref{App:fluid}, the dressed propagator reads after straightforward calculation
{\small  
\begin{align}
\label{kzint}
G^{dressed} = &\left( \frac{8 \pi G \gamma^2}{-(\omega + i \epsilon)^2 + \vec k^2}\right)^2\frac{f_{\rm sound}(\omega, \vec k)}{-\omega^2  + c_s^2 k^2 - i k^2 \Gamma_s \omega}\Big|_{\omega= vk^z}  
+  \left( \frac{8 \pi G \gamma^2}{-(\omega + i \epsilon)^2 + \vec k^2}\right)^2\frac{f_{\rm shear}(\omega,\vec k)}{\omega + i \eta_s k^2 }\Big|_{\omega= vk^z} 
\end{align}
}In a viscous fluid, the sound modes are attenuated by diffusion and as a
consequence the pole of the phonon propagator moves into the negative complex
plane; $\Gamma_s k^2$ is the sound attenuation length with $\Gamma_s =
(\frac{4}{3}\eta + \zeta)/(\rho + P) \sim l_{\rm mfp}$. In addition to the sound
mode, also long-lived, transverse shear modes can be excited in viscous fluid
whose decay is governed by specific shear viscosity $\eta_s= \eta/(P + \rho)$.
The explicit forms of the numerator structures $f_{\rm sound}(\omega, \vec k)$
and $f_{\rm shear}( \omega,\vec k)$ that are rational functions in $k_\perp$, $k_z$,
and in $\omega$ are relegated to Appendix~\ref{RS}.

The terms with viscosity are easily computed by completing the contour in the
upper complex plane. As the hydrodynamic expression is valid up to linear order
in $\eta_s$ and $\Gamma_s$, we may expand the integrand to linear order in the
viscosities. Then the only non-analytic features of the integrand arising from
the sound channel on the upper complex half-plane are just given by the graviton
pole $k_{grav}^z = - \frac{k_\perp}{\sqrt{v^2 -1}}$ and another spurious pole given by
$k_*^z = i k_\perp$. In the case $v< c_s$, there is an additional pole of the
hydrodynamic propagator that is on the upper complex half plane $k_*^z = i \frac{c_s}{\sqrt{c_s^2 - v^2}}$ and the
contribution simply reads

\begin{align}
\label{sound}
{\rm Im}\int \frac{dk_z}{2\pi} & k_z \left( \frac{1}{-(\omega + i \epsilon)^2 + \vec k^2}\right)^2   \frac{f_{\rm sound}(\omega, \vec k)}{-\omega^2 + c_s^2 k^2- i k^2 \Gamma_s \omega } 
=(P+\rho)\frac{ \left(v^2+1\right)^2}{4 v^2 k_\perp^2} \theta(v-c_s)\\
&+(P+\rho)\frac{v \left(v^2+1\right)^2 }{8 c_s k_\perp \left(c_s^2-v^2\right)^{3/2}}\theta(c_s-v)\Gamma_s  \nonumber \\
&+ 
\frac{(P + \rho)}{k_\perp v} \Bigg\{
\frac{(v^2-4)   \sqrt{1-v^2}-4 v^2  }{2 }
+ 
\theta(c_s-v) 2\frac{c_s \left(v^2+1\right) }{ \sqrt{c_s^2-v^2}}
\Bigg\}\eta_s,
\end{align}
where the first term corresponds to the ideal fluid dynamics and
\begin{align}
\label{shear}
{\rm Im} \int \frac{dk_z}{2\pi} k_z \left( \frac{1}{-(\omega + i \epsilon)^2 + \vec k^2}\right)^2\frac{f_{\rm shear}(\omega, \vec k)}{\omega + i \eta_s k^2 }  =  2\frac{v (P+\rho)}{ k_\perp}\eta_s
\end{align}
We show the resulting drag force of the viscous fluid on the bullet and compare it to the ideal 
liquid behavior in Fig~\ref{fig:viscosity_all} where the red dashed lines correspond to the above expansion. As expected, the force does not vanish in the subsonic regime, where the 
viscous term is the proper leading order calculation of the force.

\subsubsection{Breakdown of the gradient expansion at crossing the sound barrier}
\label{sec:breakdown}
\begin{figure}[t]
	\includegraphics[width = 0.5\textwidth]{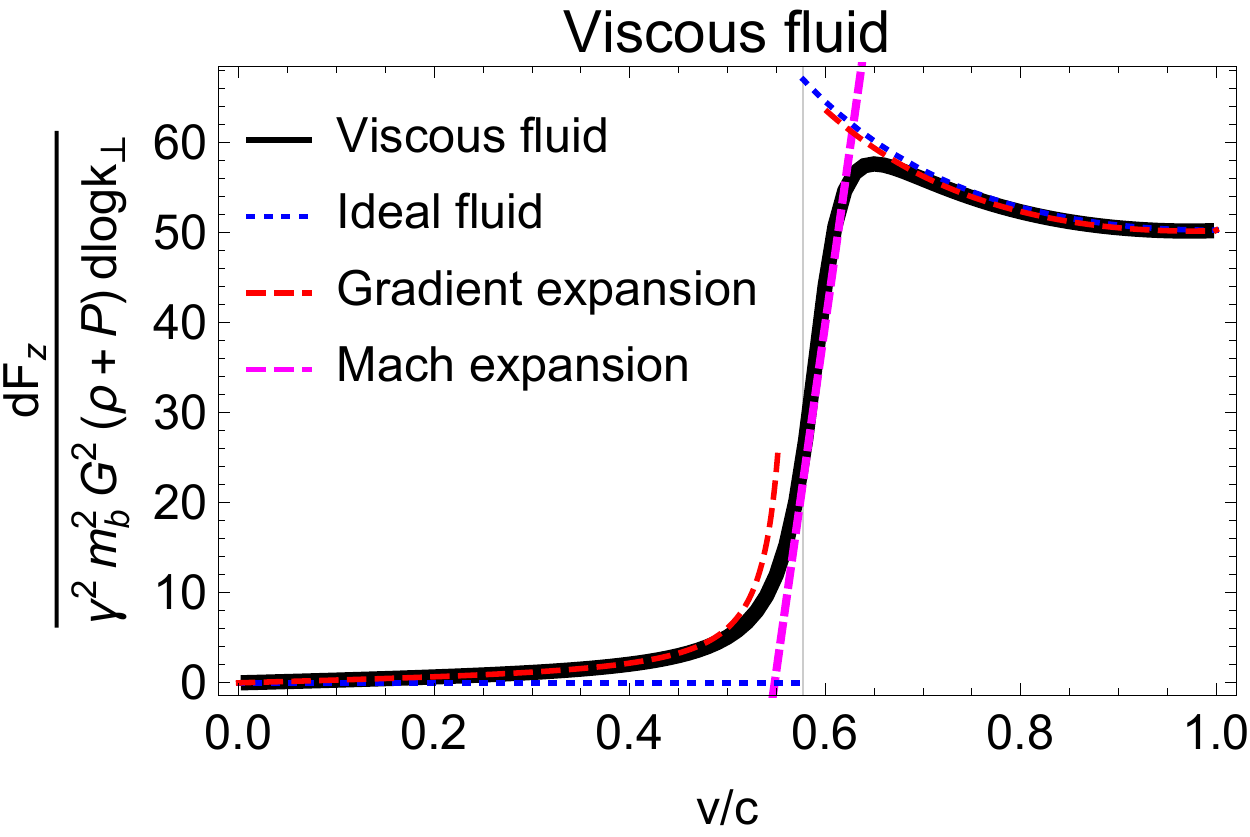}
	\caption{The dynamical friction in realistic viscous fluids. For purposes of illustration we show the drag force for conformal fluid for which $c_s^2=1/3$ (marked by a vertical line) and $\zeta = 0$ and display the contribution arising from scale $k_\perp \Gamma_s = 0.01$. The thick black line corresponds to the drag force in viscous fluid computed directly from the correlation function of (\ref{kzint}) without further expansions. This full result is compared to ideal hydrodynamics (blue dashed line, Eq.~(\ref{ideal_fluid})), gradient expansion of the first-order fluid dynamics (red line,  eqs.~(\ref{sound}) (\ref{shear})). The gradient expansion fails for velocities near the speed of sound  $|v-c_s| < \textstyle{\frac{(\Gamma_s k_\perp)^{2/3}}{2 c_s }}$. Within this window, the ideal hydrodynamic expression receives $\mathcal{O}(1)$ viscous corrections irrespective of the magnitude of viscosities $\eta$ and $\zeta$. Within this window the drag is well described by the Mach expansion (purple dashed line, Eq.~\ref{Mach}).}
	\label{fig:viscosity_all}
\end{figure}

If the velocity of the projectile approaches the speed of sound from above,
according to~\eqref{openingangle} the direction of the emitted phonons and the
bullet coincide. In this case, the energy is not transmitted away from the
projectile but the sonic boom created by the projectile follows its trajectory,
see for example~\cite{Ostriker:1998fa} for an illustration. Because the phonon travels with the
bullet for an arbitrarily long time, it will become sensitive to viscous
corrections for arbitrary small $k_\perp$. That this is the case is seen explicitly in (\ref{sound}) where 
the relative correction to the ideal fluid result 
of~\eqref{ideal_fluid} is enhanced near the Mach limit
\begin{align}
F^z = F^z\Bigg |_{ideal fluid} \left[1 + \mathcal{O}\left(\frac{k_\perp \Gamma_s }{(c_s^2 -v^2)^{3/2}}\right)\right].
\end{align}
This shows that for velocities in the window $|(v^2-c_s^2)| < (k_\perp \Gamma_s)^{2/3}$ the first viscous corrections give a contribution comparable to the leading order correction and the leading order fluid-dynamical picture does not give correct description at any $k_\perp$ within this $k_\perp$-dependent window. 

To clarify the source of this behavior, consider a bullet moving exactly at the speed of sound $v=c_s$. In this case, the leading order $k^z$-dependent part of the denominator sound propagator vanishes exactly and the unexpanded propagator reads 
\begin{align}
\label{csprop}
\frac{1}{-\omega^2 + c_s^2 k^2- i k^2 \Gamma_s   \omega}\overset{v= c_s}{\longrightarrow} \frac{1}{c_s^2 k_\perp^2 - i c_s k^2 k^z \Gamma_s}.
\end{align}
The denominator is a cubic function of $k^z$ and has three poles, one in the upper complex plane on the imaginary axis and two in the lower complex half-plane.%
 \footnote{The two poles in the lower complex half-plane become the ideal hydrodynamic poles in the ideal limit at supersonic velocities. In the subsonic case, these two poles merge in the ideal limit to form the single pole in the lower complex half-plane of ideal fluid-dynamic propagator. The pole in the upper complex half plane reduces to the ideal one in the subsonic region and becomes irrelevant in the supersonic regime. }
The location of the pole in the upper complex half-plane is inversely proportional to $\Gamma_s$ and found straightforwardly by setting $k_\perp \ll k_z$ in (\ref{csprop})
\begin{align}
\label{crazy_pole}
k_{\Gamma_s}^z = i \frac{c_s^{1/3} (k_\perp\Gamma_s)^{2/3}}{\Gamma_s},
\end{align}
see Fig. \ref{poles} for illustration. That is, upon approaching the Mach limit
from below, the ideal-fluid-dynamic pole in the upper complex plane moves
towards $i$-infinity, but the growth of the imaginary part is regulated by the
viscosity and the pole goes far away from the real axis but remains in the
finite complex plane. If we had expanded the denominator before taking the Mach
limit, this pole would have not been present.

The contribution arising from this new pole---absent in ideal fluid dynamics---in the Mach limit
 is simply found by computing the integral on left hand side of equation (\ref{sound}), again completing the contour in the upper complex half-plane. The contribution arising from the pole of (\ref{crazy_pole}) gives
\begin{align}
\label{Mach}
{\rm Res}\Bigg[ k_z \left( \frac{1}{-\omega^2 + \vec k^2}\right)^2  & \frac{f_{\rm sound}(\omega, \vec k)}{-\omega^2 + c_s^2 k^2- i k^2 \Gamma_s \omega }, k_{\Gamma_s}^z \Bigg] =  \\ \nonumber
& - (P+\rho)\frac{(1+c_s^2)^2}{6 c_s^2 k_\perp^2} \left[ 1- \frac{2}{3 c_s^{1/3}} \frac{v-c_s}{(k_\perp \Gamma_s)^{2/3}}+ \mathcal{O}(k_\perp \Gamma_s) + \mathcal{O}\left( \left(  \frac{v-c_s}{(k_\perp \Gamma_s)^{2/3}} \right)^2\right)\right]
\end{align}
It is noteworthy that in the first term, the factors of $1/\Gamma_s$ cancel exactly between the numerator and the denominator, such that the contribution is independent of $\Gamma_s$ and remarkably survives even in the limit of ideal fluid $\Gamma_s \rightarrow 0$. That is, the fluid-dynamical expansion breaks down in the sense that first-order viscous corrections become as large as the leading order ideal fluid result near the Mach limit. However, the fluid-dynamical expansion still remains a good expansion in the sense that the corrections arising from second-order fluid dynamics---\emph{i.e.} second order corrections to Eq.~\ref{hydro}---still remain subleading compared to the full leading order result. 

We finally note that the integral over $k^z$ in (\ref{kzint}) is straightforwardly performed analytically by computing the residues of the poles in the upper complex half-plane for any $v\neq c_s$ but leads to a complicated expression that we have decided not to include here (for a similar calculation in a different context, see \cite{Neufeld:2008fi}). 
Fig.~\ref{fig:viscosity_all} displays the drag force in conformal viscous fluid (with $c_s^2=1/3$ and $\zeta=0$). The thick black line corresponds to evaluating Eq.~\ref{kzint} without making further expansions, whereas the red and purple dashed lines correspond to the gradient expansion of Eqs.~(\ref{sound}) and (\ref{shear}) and the Mach expansion of (\ref{Mach}), respectively.

\subsection{Free streaming limit}
We now discuss the drag force in the opposite limit of non-interacting medium. We provide results for a gas whose constituents may have any kinematics, thus extending the existing results on non-relativistic and ultra-relativistic approximations appearing in the literature.

In the absence of collisions, a system of freely streaming particles in a background metric $g^{\mu\nu}$ is described by transport equation
\begin{align}
p^\mu \partial_\mu f(t, \vec x; \vec p) - \Gamma^\alpha_{\phantom \beta \beta  \gamma}
p^\beta p^\gamma \nabla^{(p)}_\alpha f(t, \vec x; \vec p) = 0.
\end{align}
The distribution function $f$ describes a distribution of on-shell particles
whose energies are given by a dispersion relation $p^0 = E(\vec p) = \sqrt{	|\vec p|^2 + m_g^2}$ and the partial $p^0$-derivative vanishes, $
\nabla^{(p)}_0 f = 0$. Here $m_g$ stands for the mass of the particles in
the gas.

Assuming a small correction in the distribution function $\delta f=f-f_0$ due to the metric perturbation $h_{\alpha \beta}$,
this equation is solved in Fourier space 
\beq \label{branchcut}
\delta f = \frac{-i}{p^0} \frac{\Gamma^i_{\phantom \beta \beta \gamma }
	p^\beta p^\gamma
	\nabla_i^{(p)}f_0}{- (\omega+ i \epsilon) +  \vec v_p \cdot \vec k }.
\eeq  
Here we have introduced the velocity $ \vec v_p  = \vec p/p^0$ of the medium
constituent particles (not to be confused with the velocity of bullet
in~\eqref{eq:tmunubullet}). The numerator reflects the coupling of the medium
constituent particles with the gravitational field and the denominator is the
retarded eikonal propagator of the free-streaming particles; in coordinate space
it reads $\sim\delta(\vec{x} - \vec v_p t)\theta(t)$. With this solution at hand
it is simple to solve for the Green function of the energy-momentum tensor
\beq
\label{eq:dTmunu}
G^{\mu \nu, \alpha \beta }_{kin} = \frac{\delta T^{\mu \nu}}{\delta h_{\alpha \beta }} \ \ \ \ {\rm with } \ \ \ \ 
\delta T^{\mu \nu } = \int \frac{d^3p}{(2\pi)^3} \frac{p^\mu p^\nu}{p^0} \delta f
\eeq 
where the dependence on $h_{\alpha \beta}$ comes from the Christoffel symbols. 

The dressed graviton propagator~\eqref{eq:Gdressed} can then simply be written with the aid of the solution 
$\delta f$~\eqref{branchcut}, the definition of the energy-momentum Green function~\eqref{eq:dTmunu}, 
and the graviton propagator~\eqref{eq:gravition} 
{\small
\beq
\label{Gdres}
G^{dressed} = -i \int \frac{d^3p}{(2\pi)^3 } \left(
\frac{\kappa }{-(\omega + i \epsilon)^2 + k^2 }  
\right)^2 I_{\gamma \delta, \mu \nu} I_{\alpha \beta, \rho \sigma}
\frac{\delta \Gamma^i_{\phantom{\alpha} \omega \xi}}{\delta h_{\rho \sigma }}
\frac{p_i p^\mu p^\nu p^\omega p^\xi}{(p^0)^2}
\frac{f_0'(p^0)}{- (\omega+i \epsilon) + \vec k \cdot  \vec v_p} u^\gamma u^\delta
u^\alpha u^\beta
\eeq  
}%
which we again study for $\omega = v k_z$. A similar result is found in, \emph{e.g.},~\cite{Rebhan:1990yr}. Here we have also assumed
that the unperturbed distribution function is isotropic, such that it only
depends on the energies of particles $p^0$ and we may write $ \nabla^{(p)}_i f_0
= \frac{p_i}{p^0} f_0'(p^0)$---anisotropic systems exhibit interesting
dynamics that go beyond the scope of this work (see
\emph{e.g.}~\cite{Romatschke:2003ms, Kurkela:2011ti} for complex dynamics in 
anisotropic non-Abelian plasmas). 
The projection tensors
arising from graviton propagators $I_{\gamma\delta,\mu \nu}$ are defined in the
Appendix~\ref{App:graviton}. Expressing the integral over particle momenta in
spherical coordinates $d^3p =p^2 dp d\Omega_p$ allows to perform the integral over
the direction of the particles $d\Omega_p$
\beq\label{Gdressedfreestream}
{\rm Im} \int k_z G^{dressed} dk_z & = & {\rm Im} \int k_z dk_z \int dp 
\frac{f'(p^0) p \gamma^4}{240 \pi^2 (k_\perp^2 + k_z^2)^{4} (p^0)^2 
	(k_\perp^2 - k_z^2 (-1+v^2))^2} \times \\ \nonumber &&
\left( 
R(k_z, k_\perp; p, p^0; v )+ S(k_z, k_\perp; p, p^0; v) \log 
\frac{k_z p^0 v + \sqrt{k_z^2 + k_\perp^2} p}{k_z p^0 v - \sqrt{k_z^2 + k_\perp^2}p} 
\right)  
\eeq 
where functions $R$ and $S$ arise from the numerator algebra in (\ref{Gdres}) and are fully analytic in all their variables---for the explicit form see Appendix~\ref{RS}.

In order to perform the final integral over $k_z$ it is again useful to consider
the analytic structure of the integrand~\eqref{Gdres}. Again, the integrand has
non-analytic structures where propagators go on shell. As in the fluid-dynamic
case, the graviton poles are located on the imaginary $k_z$-axis for $\omega = v
k_z$. In contrast, the eikonal propagator of the in-medium particles $(-i \omega
+ i \vec k\cdot \vec v_p)^{-1}$  of~\eqref{Gdres} has poles on the the real axis for
\begin{equation}
\label{eq:cut}
|k_z/k| < v_p/v.
\end{equation}
 Upon integrating over the angles of the velocities of the particles $d\Omega_p$,
the integral over the poles turns into the logarithm 
of~\eqref{Gdressedfreestream} which has a branch cut along the real axis, see Fig.~\ref{poles}. The integration contour
follows the real axis just above the cut  to account for the $i \epsilon$
in~\eqref{Gdres}---to compute the imaginary part, it is enough to compute the
discontinuity along the cut. Note that the function $R$ is analytic and does
not contribute to the drag force.

There are two distinct kinematic ranges that the integral over $k_z$ can have depending 
on the momenta of the particles $p$.  For particles whose velocity exceeds that of the bullet 
\beq
p > \gamma m_g v,
\eeq      
 the denominator of the logarithm
is always negative, such that the branch cut extends over the entire 
real $k_z$-axis. The discontinuity of the branch cut of the logarithm is simply $2\pi i$, and 
the rest (the $S$ functions and the prefactor) is simply integrated from minus to plus 
infinity. The final result reads 
\beq\label{eq:freestrem_highp}
{\rm Im }\int k_z G^{dressed }_{I} dk_z & = &\frac{ 64 \pi^2 \gamma^4 G^2 }{96 \pi k_\perp^2 v^2 }
\int_{\gamma m_g v}^\infty  dp\ \frac{p}{p^0} \frac{df_0}{dp^0} \times \\ \nonumber &&
\Big[ 2v \big( 3p^4 (-5 + 3v^2) + (p^0)^4 (-27 + 29v^2) -
6p^2 (p^0)^2 (-7 + 9v^2)
\big) + \\ \nonumber &&
3 (p^2 - (p^0)^2) (-1 + v^2) \Big( (p^0)^2 (9 - 5v^2) + p^2 (-5 + v^2) \Big) \log \left( \frac{1+v}{1-v}\right)
\Big].
\eeq
Using $\frac{p}{p^0}\frac{d f_0}{dp^0} =\frac{d f_0}{dp} $ and performing the partial integral, the
expression reads simply
\beq
\label{ideal_1}
{\rm Im} \int k_z G^{dressed }_{I} dk_z & = & -\frac{64 \pi^2 \gamma^4 G^2 }{12 \pi k_\perp^2 v^2 }
\int_{\gamma m_g v}^\infty  dp
\ p f_0  \times \\ \nonumber &&
\Big[  2 m_g^2 \left(v^2-3\right) v-16 p^2 v^3+3 m_g^2 \left(v^2-1\right)^2 \log \left(\frac{1+v}{1-v}\right)
\Big]
\eeq 
plus a boundary term at $p = \gamma m_g v$, which always exactly cancels with the $p<\gamma m_g v$ contribution as 
the integrand is continuous.
Note that in the limit of  ultra-relativistic medium the branch cut always extends over the entire real 
axis and therefore the entire contribution to the dynamical friction is given 
by~\eqref{eq:freestrem_highp} with the integration running from zero to infinity. 

However, in the generic case this is not the only contribution. 
While the contribution to the integral from particles that move at velocities slower than the bullet 
\beq
p < \gamma m_g v
\eeq  
is negligible in the case of ultra-relativistic particles, it becomes dominant in the 
non-relativistic limit. The branch cut is still located on the real axis, but it does not extend all the way to infinity. It is rather located between 
\beq\label{eq:kzrange}
-\frac{k_\perp p}{\sqrt{p^2 (v^2 -1) + v^2 m_g^2}} < k_z < 
\frac{k_\perp p}{\sqrt{p^2 (v^2 -1) + v^2 m_g^2}}.
\eeq  

The result of the integration in the second physical region is 
\beq\label{eq:freestrem_lowp}
\label{ideal_2}
{\rm Im} \int k_z G^{dressed}_{II} dk_z  & =& \frac{64 \pi^2 \gamma^4 G^2}{48 \pi k_\perp^2 v^2}
\int_0^{\gamma m_g v} dp\ \frac{f'_0(p^0) p}{p^0} \times \\ \nonumber &&
\Big[ -3p(p^0)^3 (9-14v^2 + 5v^4)  + p^3 p^0 (29 - 54v^2 + 9v^4) +\\ \nonumber &&
3 (p^2 - (p^0)^2)(-1 + v^2) \big( (p^0)^2 (9-5v^2) + p^2 (-5 + v^2)
\big) \arctanh \left(\frac{p}{p^0}\right)
\Big].
\eeq

\begin{figure}[t]
	\centering
	\includegraphics[width=0.47\textwidth ]{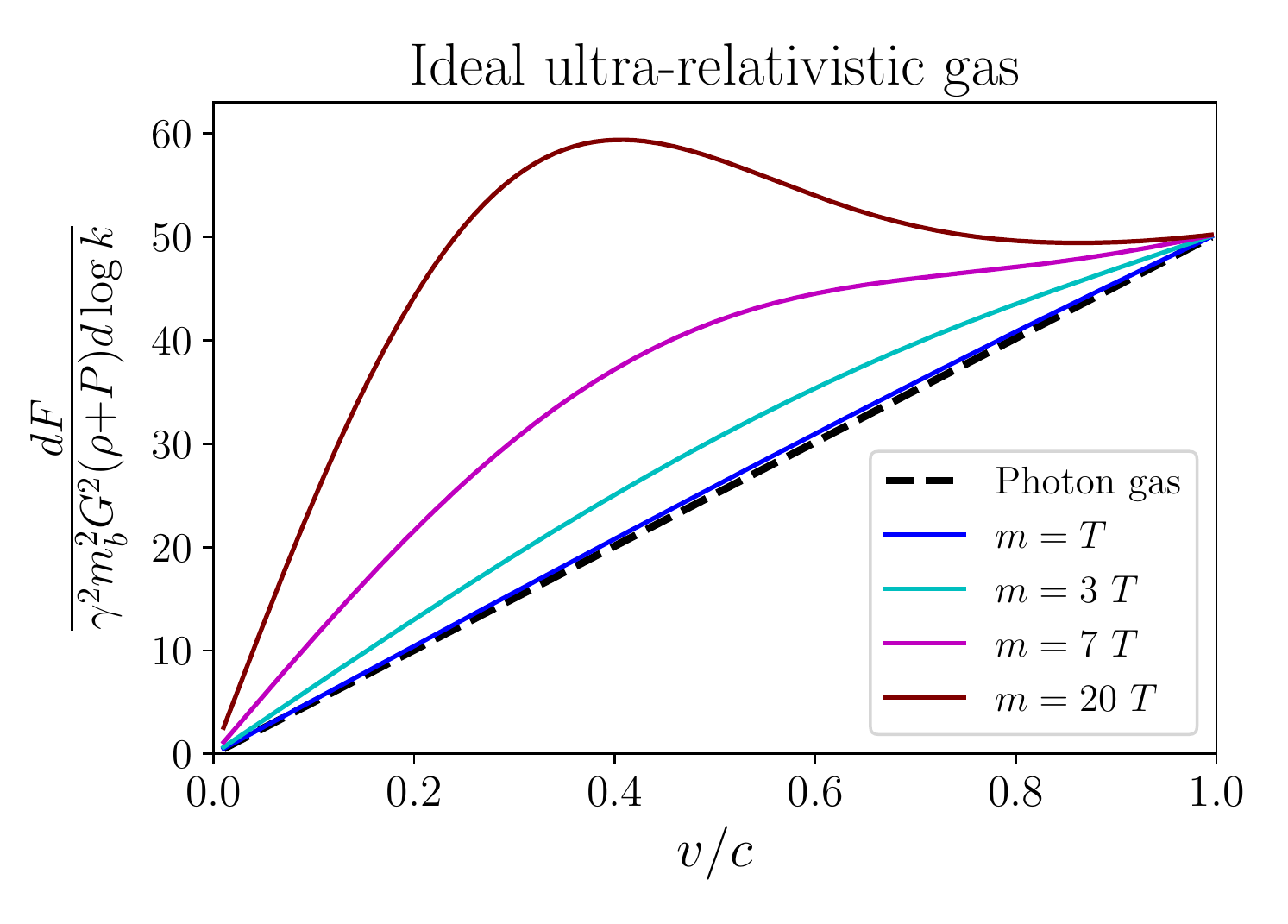}
	\includegraphics[width=0.47\textwidth ]{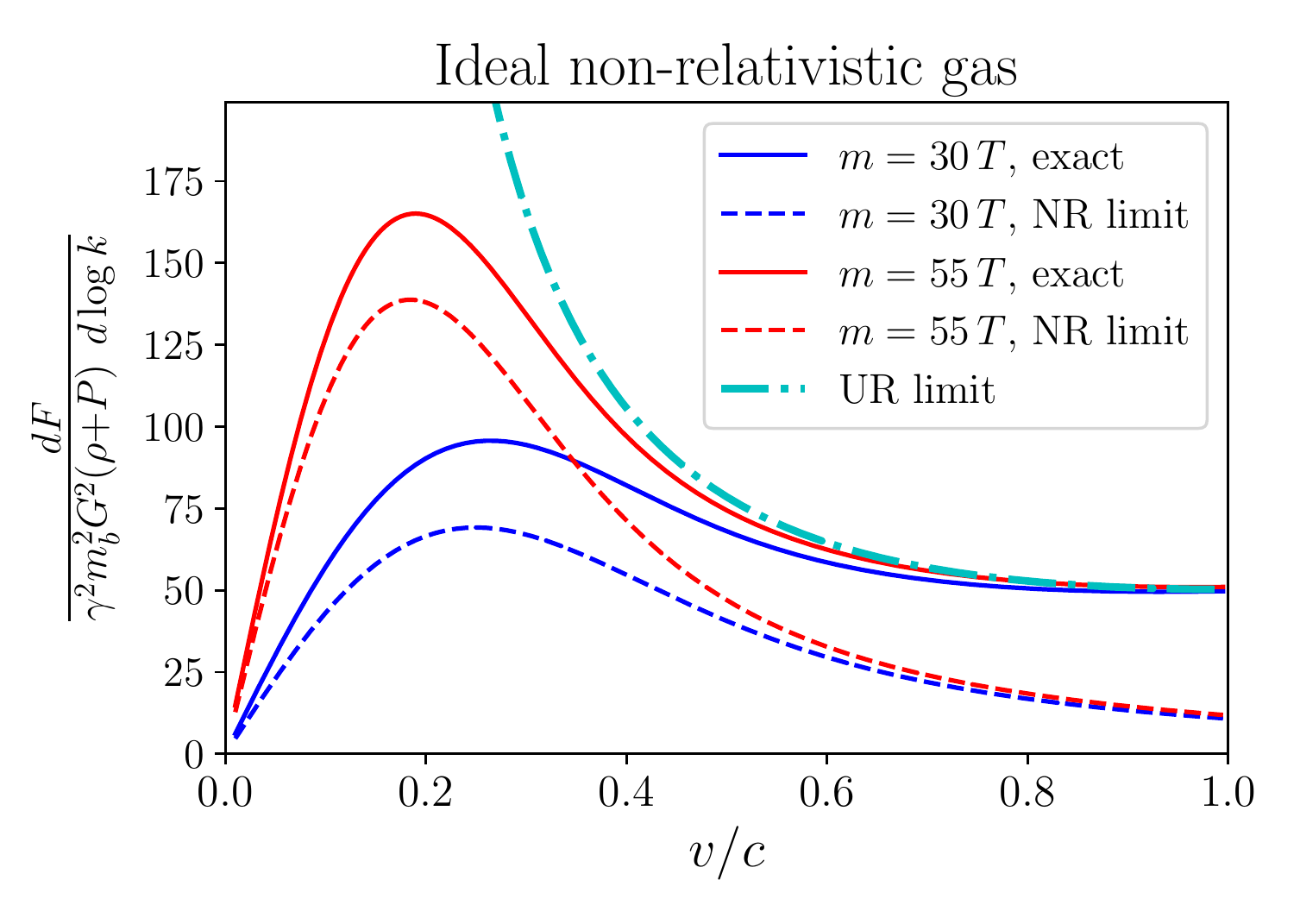}
	\caption{The dynamical friction in ideal gas. The left panel shows the force in the regime of  relativistic or
	nearly-relativistic gas, comparing the effect to the ultra-relativistic approximation.
	The right panel shows the drag force in largely
	non-relativistic gas. The solid lines show the exact result, while the dashed
	lines stand for the non-relativistic approximation of \cite{1943ApJ....97..255C}. The dashed-dotted light
	blue line stands for the approximation of the relativistic projectile in a
	non-relativistic medium, Eq.~\eqref{eq:fastbulletslowgas}.  
	}
	\label{fig:freestream}
\end{figure}

Now let us compare the expressions that we have found to the known expressions
in the literature, namely the ultra-relativistic and non-relativistic limits. The
ultra-relativistic limit is simple to reproduce: as we have already mentioned, the branch cut
extends over the entire real axis and only the dressed propagator from
Eq.~\eqref{eq:freestrem_highp} contributes, with the lower integration limit
trivially replaced by $0$. In this limit the entire second term proportional to $\log$ inside the brackets vanishes, while the first term becomes
\beq
\frac{dF^z}{d\log k_\perp} = \frac{8 m_b^2 \gamma^2 G^2}{3 \pi} v
\int_0^\infty \frac{p^5 f'_0(p)}{p} dp.
\eeq   
If one specifies the initial distribution to be the thermal distribution $f = e^{-p/T}$ for which the energy 
density is $\rho = 3T^4/\pi^2$, one obtains 
\beq\label{eq:Syer_result}
F^z = \frac{64 \pi}{3}  G^2 m_b^2  \rho v  \gamma^2 \log \Lambda~.
\eeq 
Thus we readily recovered the  result of~\cite{1994MNRAS.270..205S}, where
$\Lambda \sim R_{\rm max}/R_{\rm min}$ arises from the regulation of the
gravitational Coulomb divergence.

In the opposite limit, where both the bullet and the medium particles are moving
at non-relativistic speeds ($v \ll 1$ and $v_p \ll 1$), we may expand the integrand in powers of  $p/ m$.
We first notice that in this limit the dominant contribution to the integral over the momenta $p$ comes from 
the region~\eqref{eq:freestrem_lowp}. 

Expanding~\eqref{eq:freestrem_lowp} to leading order in $v$ and $p/m$ and taking into account properly all the numerical factors in~\eqref{eq:fzintegral}
one finds:
\beq\label{eq:Chadra_topdown}
\frac{dF^z}{d\log k_\perp} =  \frac{2}{3\pi} \frac{m_b^2 m_g^4 G^2}{v^2} \int_0^v
f'_0(v'){v'}^3 dv'  =  \frac{2}{\pi} \frac{m_b^2 m_g^4 G^2}{v^2} \int_0^v
f_0(v'){v'}^2 dv'.
\eeq 
Here we performed a change of integration variable $v' = p/m_g$ and at the last step integrated by parts.
The would-be boundary term at $p = v$ again cancels exactly against a similar term in lower limit of the integral in~\eqref{eq:freestrem_highp}.

The drag force is brought to the form of~\cite{1943ApJ....97..255C} by
specifying the non-relativistic thermal distribution function $f_0 = e^{(\mu-m)/T}
e^{-p^2/2 mT}$, for which the energy density reads $\rho_{NR} = e^{(\mu-m)/T}
\frac{m_g^{5/2} T^{3/2}}{(2\pi)^{3/2}}$. Performing the remaining integration
gives the known result
\beq\label{eq:Chandrasekhar_result}
F = \frac{(4 \pi)^2 G^2 m_b^2 m_g }{v^2} \log \Lambda \int_0^v n(v') 
 dv'~,
\eeq 
where $n(v')$ is a number density of particles at given velocity $v'$.

We note that in addition to the above limits we can express the limit of
a relativistic bullet propagating in non-relativistic gas in a simple form.

\emph{A priori}, if the bullet is relativistic the contribution 
of the integral~\eqref{eq:freestrem_highp} is not suppressed by powers of the bullet 
velocity $v$ and must also be taken into account. However, 
in the limit of the relativistic bullet
and non-relativistic medium we expect $\gamma m_g v \gg T$ and therefore for a well-behaved 
distribution the contribution of~\eqref{eq:freestrem_highp} is exponentially suppressed. 
Therefore, in order to obtain this limit one should merely recover all the velocity factors 
in~\eqref{eq:freestrem_lowp} and to carry the integration up to infinity, since this again 
will result only in exponentially suppressed corrections:
\beq\label{eq:fastbulletslowgas}
\frac{dF^z}{d \log k_\perp} = \frac{2}{\pi} \frac{m_b^2 m_g^4 G^2 \gamma^2(1 + v^2)^2}{v^2}
\int_0^\infty f_0(v') {v'}^2 dv' = \frac{4 \pi \rho m_b^2 G^2 \gamma^2 (1+v^2)^2}{v^2}~,
\eeq  
where $\rho$ is the energy density of the non-relativistic gas. We find in agreement with \cite{Ostriker:1998fa} that for a ultrarelativistic bullet, the form of the force is identical in both ideal gas and in ideal fluid. 

Of course in the fully generic case, when we keep the mass of the medium constituent particles
arbitrary, the result is more complicated and one may need to perform
the integration over both contributions~\eqref{eq:freestrem_highp}
and~\eqref{eq:freestrem_lowp} numerically. It is instructive to compare the dynamical
friction in the different kinematic ranges. In Fig.~\ref{fig:freestream} we
specify the thermal Maxwell distribution $f = e^{- p^0/T}$ and display the drag
force for different $m/T$-ratios comparing to the non-relativistic and
ultra-relativistic limits. For particles with masses $m \lesssim T$, the force is
a monotonically rising function of velocity and is well described by the
ultra-relativistic limit of (\ref{eq:Syer_result}) shown as the blue dashed line in the left panel of Fig.~\ref{fig:freestream}. As the mass is increased (or temperature
decreased), the force as a function of $v$ develops a characteristic shape of the non-relativistic limit (\ref{eq:Chandrasekhar_result}), maximized when the velocity of the projectile is comparable to the medium particles $v\sim \langle v_p \rangle$ (red and blud dashed lines in the right panel). With finite $m/T$, the limit of large $v$ coincides with the result in ideal gas (\ref{eq:fastbulletslowgas}) shown by the dash-dotted line in the right panel of Fig.~\ref{fig:freestream}.

\subsection{Interacting kinetic theory}

We now move on to compute the drag force in a full interacting kinetic theory model. The contributions to drag force arising from the two limits discussed in the previous sections---that is $k \gg 1/l_{\rm mfp}$ and $k \ll 1/l_{\rm mfp}$---are universal in the sense that they do not depend on the microscopic details of the interactions of the medium constituents. This is no longer the case for the contribution arising from the scale of the interactions $k \sim 1/l_{\rm mfp}$ which is sensitive to the specific form of the collision kernel. To study the qualitative features that are present in an interacting kinetic theory,  we will here concentrate in a particularly simple kinetic-theory model which retains qualitative features common to all interactions while still being analytically tractable. We study the kinetic theory in the \emph{relaxation-time approximation} given by
\beq
p^\mu \partial_\mu f - \Gamma^\alpha_{\phantom \beta \beta  \gamma}
p^\beta p^\gamma \nabla^{(p)}_\alpha f = 
\frac{p^\alpha u_\alpha^{\rm rest}}{\tau} (f - f_{eq})~,
\eeq 

where the collision kernel on the right hand side of the Boltzmann equation is based on the physical assumption that interactions bring the distribution function to its equilibrium form $f_{\rm eq}$ on a given relaxation timescale $\tau\sim \tau_{\rm scat} \sim l_{\rm mfp}/v_p$ in the local rest frame of the system $u_\alpha^{\rm rest}$ satisfying the Landau condition $u_{\rm rest}^\mu T^{\nu}_{\mu} = - \epsilon u^\nu_{\rm rest}$. While this model is a gross simplification of more realistic kinetic-theory models, it serves here as a prototype model to demonstrate the effect of finite interactions; in particular, it interpolates between free-streaming behavior at scales $k\tau \gg 1$ and ideal fluid behavior at scales $k\tau \ll 1$~\cite{Romatschke:2015gic}. The formalism here discussed can, of course, be applied to more complicated interactions as needed. 

\begin{figure}[t]
	\centering
	\includegraphics[width = .75\textwidth]{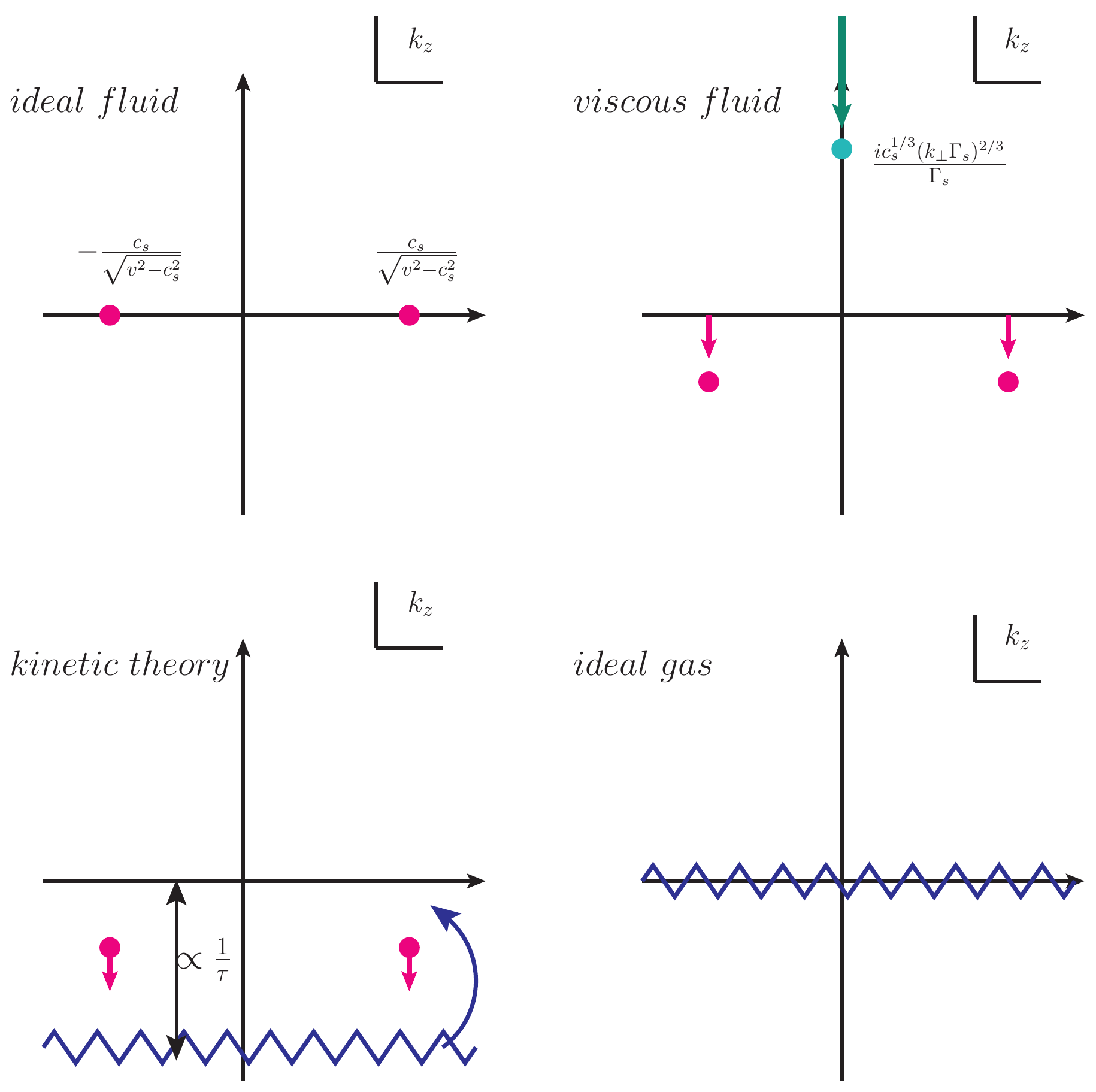}
	\caption{The analytic structure of the retarded Green function of the energy-momentum tensor $G^{\rm medium}$ in fluid dynamics and in various regimes of the interacting kinetic theory. The ideal-fluid panel shows the poles when $v>c_s$. 
The arrows in the case of the viscous fluid show the motion of the poles in the complex plain as the sound attenuation length
$\Gamma_s k^2$ is decreased. The arrows in the kinetic theory case show the motion of the poles and the quasiparticle cut as the scattering rate $\tau$ is decreased. When interpolating from fluid-dynamics to free streaming, the sound poles collide with the quasiparticle cut for $k \tau \approx 1$ and are hidden to the next Riemann sheet. In the free-streaming scenario ($\tau \rightarrow \infty$) only the quasiparticle cut remains. }
\label{poles}
\end{figure}

The retarded Green function of the energy-momentum tensor is known in the
relaxation-time approximation in the massless limit. In order to study the drag
force for systems with arbitrary masses and interaction rates, we extend these
results for finite masses. Following closely the prescription
of~\cite{Romatschke:2015gic,Kurkela:2017xis} (see also \cite{ Baym:2017xvh}) and
in close analogy to the free-streaming case, the perturbation of the
distribution function $\delta f$ caused by the gravitational perturbation
$h_{\alpha\beta}$ has the formal solution
\beq
\label{eq:df}
\delta f =-i \frac{\delta f_{eq}/\tau + \Gamma^\alpha_{\phantom \beta \beta \gamma }
	\frac{p^\beta p^\gamma}{p^0}
		\nabla_\alpha^{(p)}f_{eq}^g}{- \omega + \vec v_p \cdot \vec k - i/\tau},
\eeq  
which differs from the non-interaction result in two ways. The first difference is that the pole of the eikonal propagator moves in the negative complex plane by an amount $i /\tau$ reflecting the loss of correlation along particle trajectory because of interactions. The second difference arises because after the gravitational perturbation the local thermal equilibrium $f_{eq}(t, \vec x)= f_{eq}^g  + \delta f_{\rm eq}(t,\vec x) $ into which the interactions drive the system is not anymore same as the global thermal equilibrium $f_{eq}^g$. The local equilibrium $f_{\rm eq}(t ,\vec x)$ depends on the local value energy-momentum tensor $T^{\mu \nu}(t, \vec x)$ that in turn depends on $\delta f _{eq}(t,\vec x)$. The perturbation of the energy-momentum tensor  $\delta T^{\mu\nu}$ needed for the Green function (\ref{eq:dTmunu}) can be obtained by first taking the appropriate integral moment of (\ref{eq:df}) to obtain a closed equation for $\delta T^{\mu \nu}$
that can be then self-consistently solved for $\delta T^{\mu \nu}$. The self-consistent solution generates additional poles which correspond to the fluid-dynamical poles  for small $k \tau$, see Fig.~\ref{poles}.

Assuming a thermal Maxwell distribution $f_{eq}^g = e^{-p^0/\beta} =  e^{u^{\rm rest}_\mu p^\mu/\beta}$, the perturbation of the local equilibrium distribution becomes
\beq
\delta f_{eq} = f_{\rm eq}^g(p^0) \frac{p^0}{T} \left( \vec v_p \cdot \delta \vec u^{\rm rest} + \frac{\delta T}{T} 
\right),
\eeq 
where $ \delta \vec u$ is the perturbation of the local rest frame $u^{rest}(\tau ,\vec x)$ determined by the Landau condition. $\delta T$ is the perturbation in the temperature $T(\tau, \vec x)$ to which the interactions drive system locally, given by the local energy density
\beq
\label{deltas1}
\delta \rho  & = & \delta T^{00} \ \ \ \ \ {\rm with} \ \ \ \ \ \delta \rho = \frac{\partial \rho}{\partial T} \delta T \\
\delta u^i & = &   \frac{\delta T^{0i}}{\rho + P}  \ \ \ \ \ {\rm with} \ \ \ \ \ \rho = \int \frac{d^3p}{(2\pi)^3} f_0(p) p^0(p)  .
\label{deltas2}
 \eeq

Then taking the appropriate integral moment of~\eqref{eq:df} together with equations~\eqref{deltas1} 
and~\eqref{deltas2} 
gives a closed set of equations for $\delta T^{\mu\nu}$
 \beq
\label{eq:dTmunu2}
\delta T^{\mu\nu} =\frac{-i}{T} \int \frac{d^3 p}{(2\pi)^3} \frac{p^\mu p^\nu}{p^0}\frac{p^0 \left( \vec v_p \cdot \vec{\delta u} + \frac{\delta T}{T} 
\right)/\tau - p^0\Gamma^i_{\phantom \beta \beta \gamma }
	v^\beta v^\gamma
		v_i}{- \omega + \vec v_p \cdot \vec k - i/\tau} f_{\rm eq}^g(p^0),
\eeq  
The angular integral in this expression is simply performed similarly to the free-streaming case. 
The integral over the  $p$ is easily done in the massless case $m_g=0$, but  we have not found an analytic solution for arbitrary masses which we will solve numerically in the following. 

 Solving the above set of equations self-consistently with the definition (\ref{eq:dTmunu}) gives the full $G^{\rm medium}$ in the interacting kinetic theory. 
With this full solution at hand, we may simply read off the hydrodynamical coefficients that describe the fluid-dynamic properties of this kinetic theory model. In particular, the shear and bulk viscosities are extracted using Kubo relations 
 \beq\label{eq:Kuboeta}
 \eta & = &  \lim\limits_{\omega \to 0} \frac{1}{\omega } G^{xy,xy}(\omega, \vec k=0) \\ \label{eq:Kubozeta}
 \zeta & = & \frac{2}{9} \lim\limits_{\omega \to 0} \frac{1}{\omega} G^{ii,jj{\tiny }}(\omega, \vec k=0),
 \eeq 
 which---together with the speed of sound $c_s^2= \frac{dP}{d\rho}$---fix the form of the correlation function in the large-wavelength limit. See Appendix \ref{hydro-coeff} for the explicit integral expression of $\eta$ and $\zeta$. However, having the full correlation function at hand, we may go beyond the fluid-dynamic approximation.
 
After solving $G_{\rm medium}$, we can again proceed to compute to drag force by contracting the medium propagator with the the graviton propagators and integrating over $k_{\perp}$ and $k_z$, as per (\ref{eq:fzintegral}). Both ideal fluid and ideal gas are dissipationless, and both the sound modes in fluid and the free-streaming particles in gas are asymptotic states of the corresponding limits. This was reflected in that the non-analytic structures of $G_{\rm medium}$  were located on the real $\omega$-axis in both cases. This is no longer the case at finite $\tau$,
where the finite interaction with the medium destroys any long-distance correlation and both the fluid-dynamic poles and the free-streaming cut move to negative complex half-plane of $\omega$; the interplay between the hydrodynamical poles and the quasiparticle-cut is discussed in detail in \cite{Romatschke:2015gic, Kurkela:2017xis}, see Fig.~\ref{poles}.

\begin{figure}[t]
	\hspace{-4mm}
	\begin{minipage}{0.5\linewidth}
		\includegraphics[width = \textwidth]{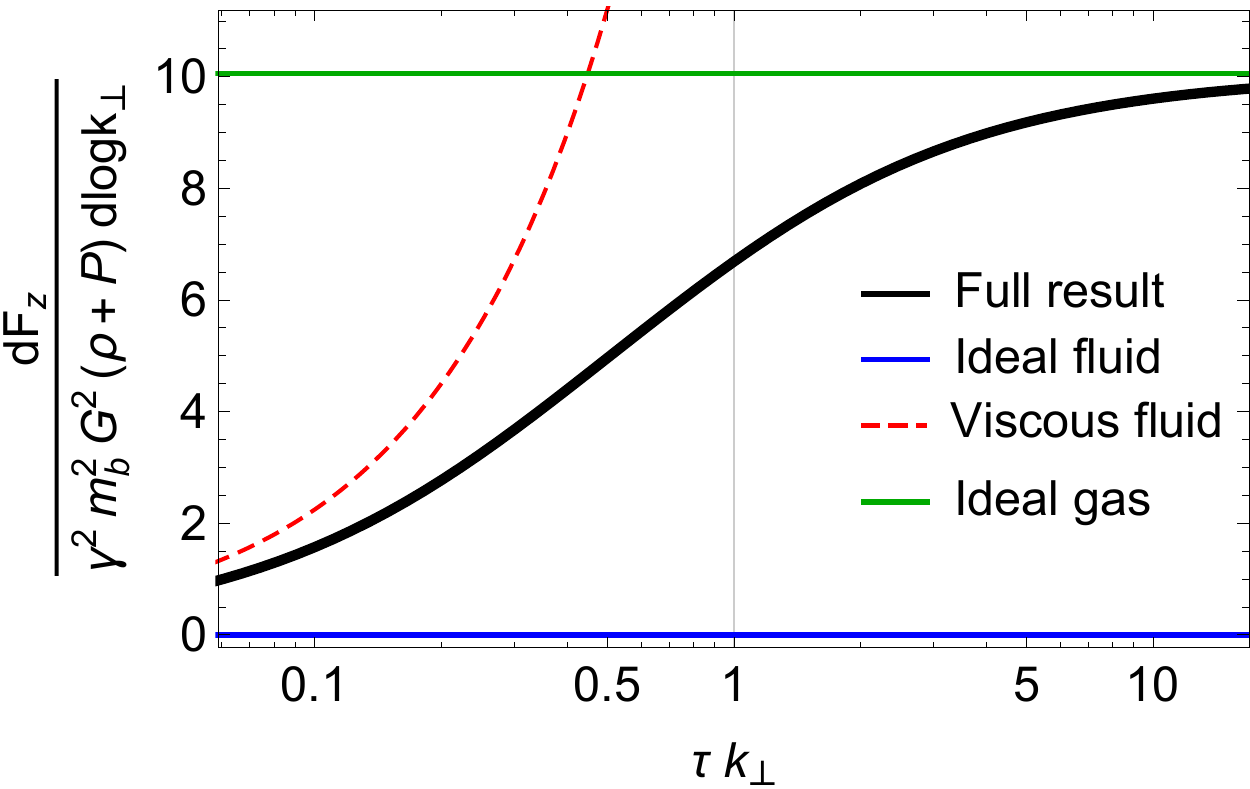}\\
		(a)  $c_s>v=0.2$.
	\end{minipage}
	\begin{minipage}{0.5\linewidth}
		\includegraphics[width = \textwidth]{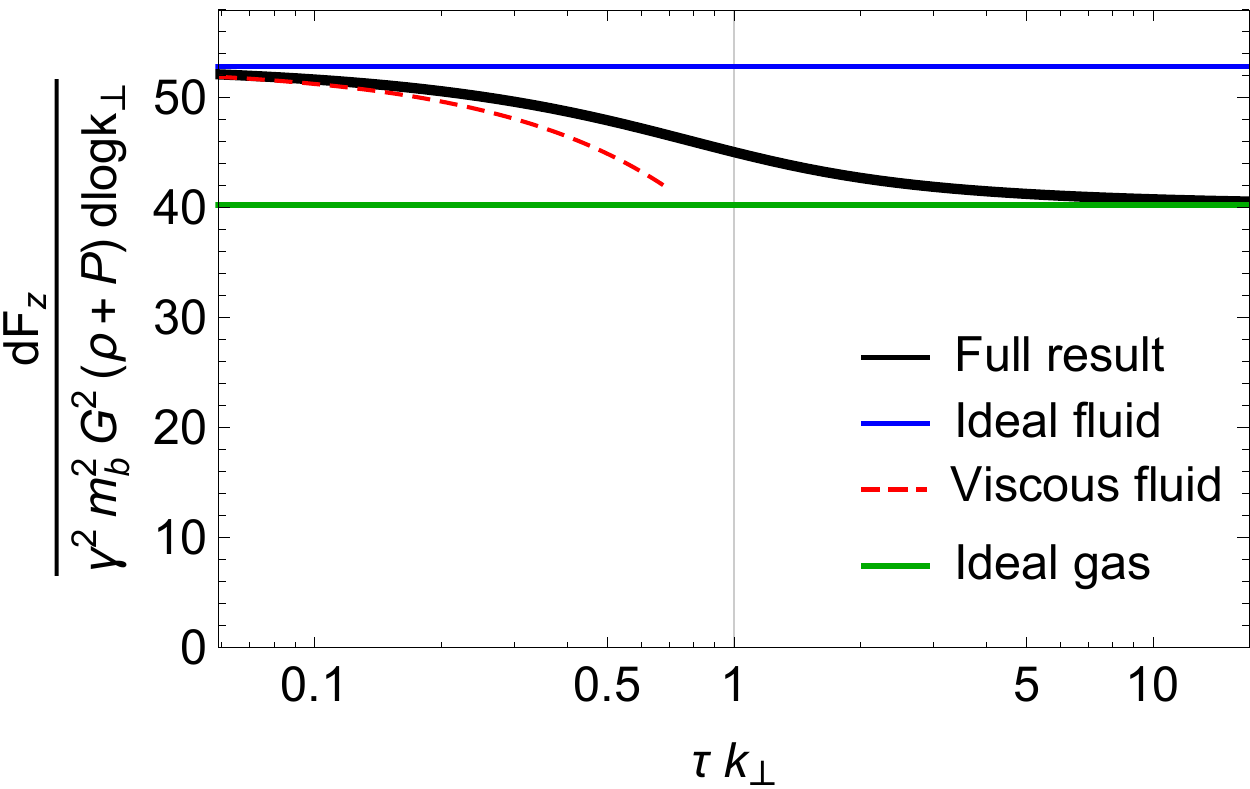}\\
		(b) $c_s<v=0.8$.
	\end{minipage}
	\caption{Differential drag force $dF^z/\log k_\perp $ as a function of $\tau
	k_\perp$ for a massless gas in interacting kinetic theory in relaxation-time approximation with $c_s^2 =1/3$. At distance scales much shorter than the mean free path $k_\perp \tau \gg 1$, the full results asymptotes to the ideal-gas result (green line). 
	 At distance scales much longer than the mean free path
	$k_\perp \tau \ll 1$, the medium exhibits ideal-fluid behavior as demonstrated
	by the agreement of the full result (black line) with the ideal fluid-dynamic
	approximation (blue line). The approach to ideal fluid is described by the first-order viscous fluid dynamics shown by the red dashed line. }
	\label{fig:viscosity}
\end{figure}

As the Green function of the energy-momentum tensor of the interacting kinetic theory is well approximated by ideal-fluid and ideal-gas limits at large and short distance scales, the contribution to the drag force arising from these scales also corresponds to the respective limits. This is demonstrated in Fig.~\ref{fig:viscosity}, where we show the differential contribution to the force arising from different scales $\textstyle{\frac{ dF^z}{d \log k_\perp}}$.
As expected, for small $k_\perp$ the full kinetic-theory contribution (black line) approaches the ideal-fluid result (blue line, Eq.~\eqref{ideal_fluid}) and the first corrections in powers of $\tau k_\perp $ to the ideal-fluid result is captured by the viscous correction of (\ref{sound}) and (\ref{shear}). Note that for $v<c_s$, (left panel), the drag force vanishes in ideal fluid, but already the first viscous correction leads to finite correction. In the opposite limit of large $k_\perp$, the full result approaches the ideal gas value. For a subsonic projectile, the interacting kinetic theory approaches the ideal gas limit from below, while for a supersonic projectile the ordering is reversed. 
 
\begin{figure}[t]
\includegraphics[width=0.48\textwidth]{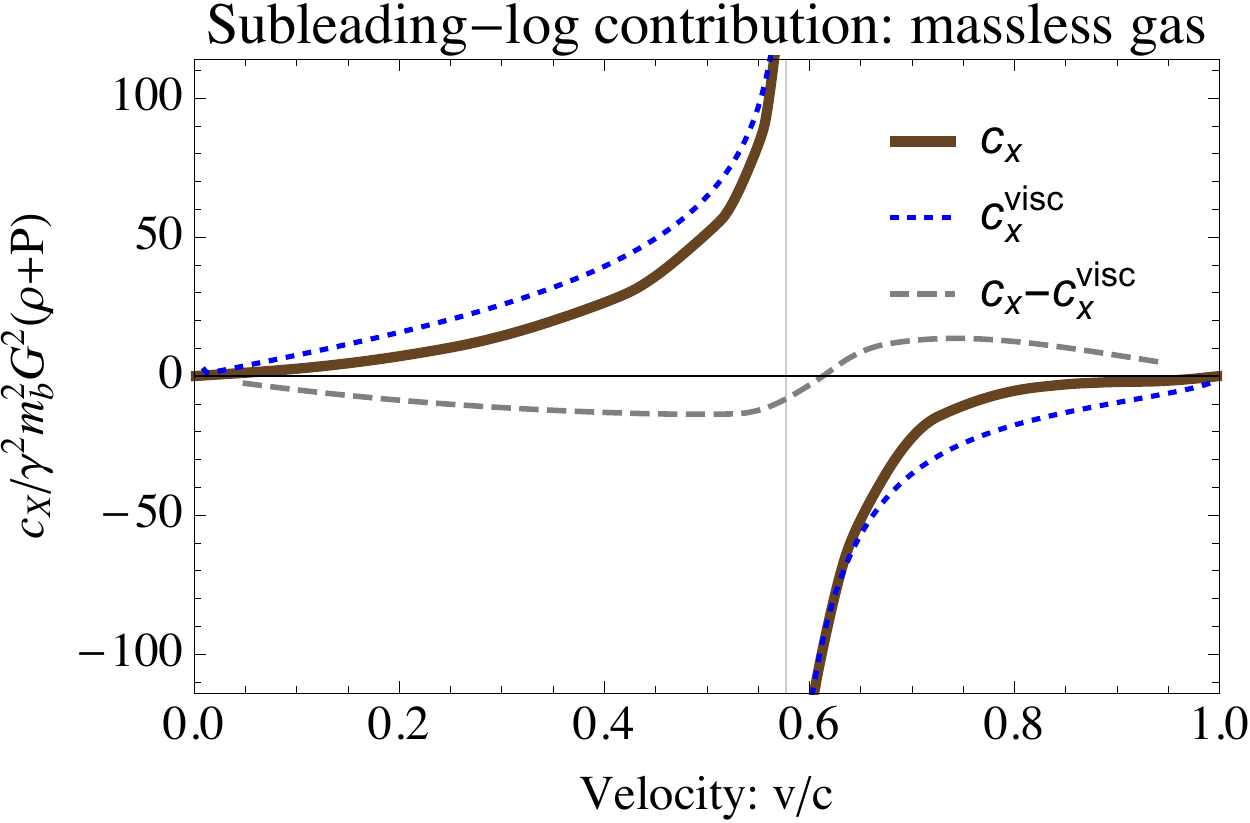}
\includegraphics[width=0.48\textwidth]{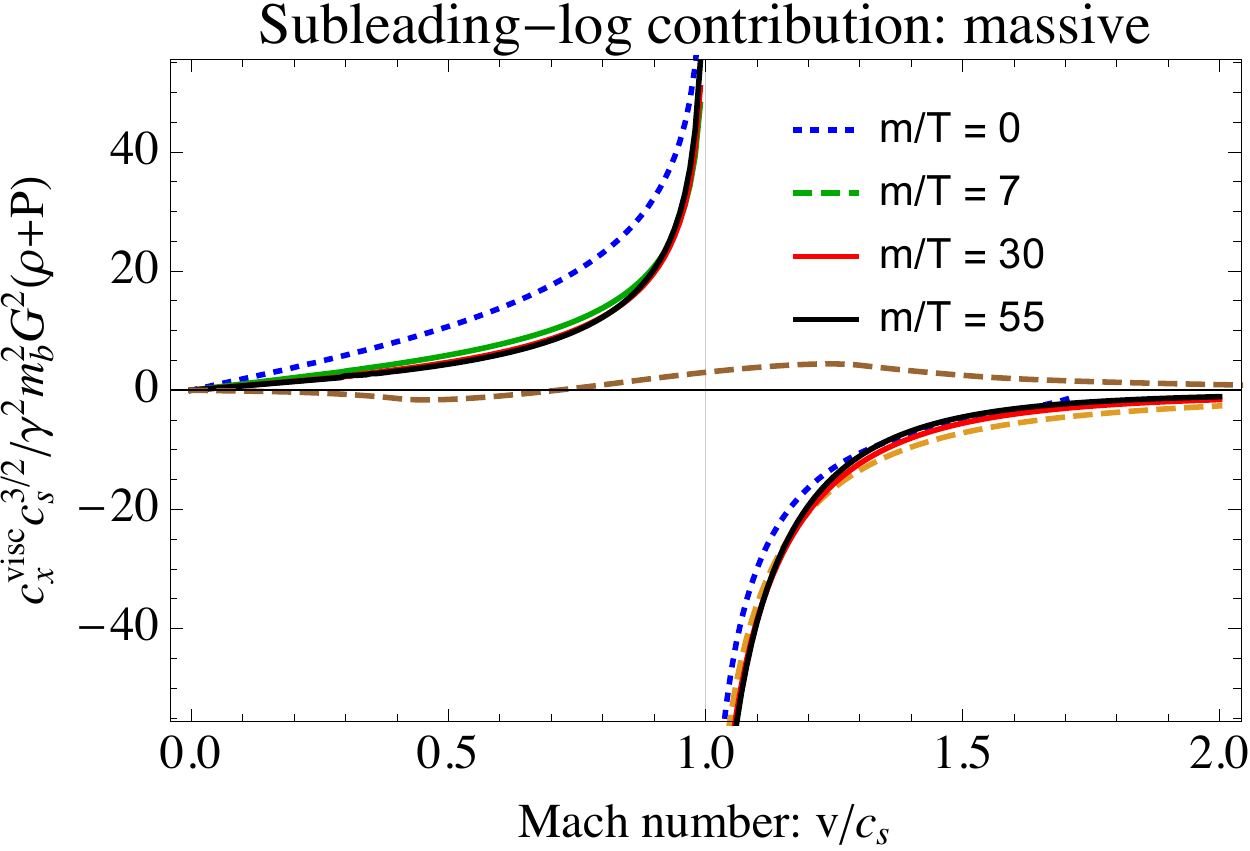}
\caption{ The sub-leading-log contribution to the drag force. Left panel shows the 
contribution $c_X$ computed in kinetic theory in relaxation-time approximation (thick brown line), and the corresponding quantity $c_X^{visc}$ in first-order, viscous fluid dynamics (blue dotted line), as defined in Eq.~(\ref{cxvisc}). The gray dashed line corresponds to the difference to the two other lines, showing that the contribution arising from the $l_{mfp}$-scale is numerically small and regular. The right panel shows $c^{visc}_{X}$ in non-confromal fluid dynamics corresponding to different $m/T$ ratios. The dashed brown line shows $c_X-c_{X}^{visc}$ computed in the full interacting kinetic theory for $m/T=7$ showing similar behavior as in massless case. }
\label{masslessPlot}
\end{figure}

In order to compute the total force, the differential force $\textstyle{\frac{ dF^z}{d \log k_\perp}}$ needs to be integrated over $\log k_\perp$ imposing a short- and a long-distance regulators to regulate the logarithmic Coulomb divergences. To provide the result in a way that does not depend on the regulators, we compute the difference between the full drag force and its leading-log expression
\begin{align}
c_X = \int_{-\infty}^{\infty} d \log k_\perp \left[ \frac{d F_z}{d \log k_\perp }-\theta(1-\tau k_\perp  )  \frac{d F_z}{d \log k_\perp }\Bigg|_\textrm{ideal fluid}-\theta(\tau k_\perp -1 )  \frac{d F_z}{d \log k_\perp }\Bigg|_\textrm{ideal gas}\right],
\end{align}
where the expression form ideal fluid is given by (\ref{ideal_fluid}) and for ideal gas in (\ref{ideal_1}) and (\ref{ideal_2}). This corresponds to the integral of difference between the full result and a step function switching from the ideal fluid to ideal gas at $\tau k_\perp = 1$ marked as a vertical line in Fig.~\ref{fig:viscosity}. This expression is free of Coulomb divergences (for $v\neq c_s$) as the integrand vanishes by construction for large and small $k_\perp$. We plot this quantity for medium of massless particles in Fig.~\ref{masslessPlot} where we see that the subleading-log corrections are largest for $v \sim c_s$ where the correction diverges and changes sign. 

The origin of this divergent behavior lies in the breaking of the fluid-dynamic gradient expansion near $v \approx c_s$ discussed in Section \ref{sec:breakdown}. The fact that near the Mach limit the drag force is never well described by ideal fluid dynamics is reflected in the fact that the remainder $c_X$ obtains a large contribution down to $k_\perp \tau \sim (c_s^2 - v^2)^{1/3}$, (see Eq.~(\ref{Mach})). To demonstrate this explicitly we also show the difference between the drag in viscous fluid compared to the ideal fluid for $k_\perp < 1/\tau$
\begin{align}
\label{cxvisc}
c^{\rm visc}_X = \int_{-\infty}^{-\log \tau} d \log k_\perp \left[ \frac{d F_z}{d \log k_\perp }\Bigg|_\textrm{viscous fluid} - \frac{d F_z}{d \log k_\perp }\Bigg|_\textrm{ideal fluid}\right].
\end{align}
This is shown as the red dashed line in the left panel of Fig.~\ref{masslessPlot}. The divergent feature of $c_X$ is contained in  $c_X^{\rm visc}$ demonstrating that the breakdown of the leading-log expression arises from long-wavelength region where viscous fluid-dynamics still can be used even if the ideal hydrodynamic treatment fails.  The remainder arising from the scale $1/\tau$, $c_X-c_X^{\rm visc}$, is indeed regular and numerically small at all $v$. The right panel of Fig.~\ref{masslessPlot} shows $c_X^{\rm visc}$ for different masses of the medium constituent masses as function of Mach number $v/c_s$. We observe that by normalizing the contribution with $c_s^{3/2}$ motivated by Eq.~(\ref{Mach}), the different lines corresponding to different approximately collapse to single universal curve for $m\gg T$. The figure also displays displays the remainder $c_X-c_X^{\rm visc}$ for $m/T=7$, which in analogy to the massless case is regular and numerically small. 
 
\begin{figure}
\includegraphics[width=0.49\textwidth]{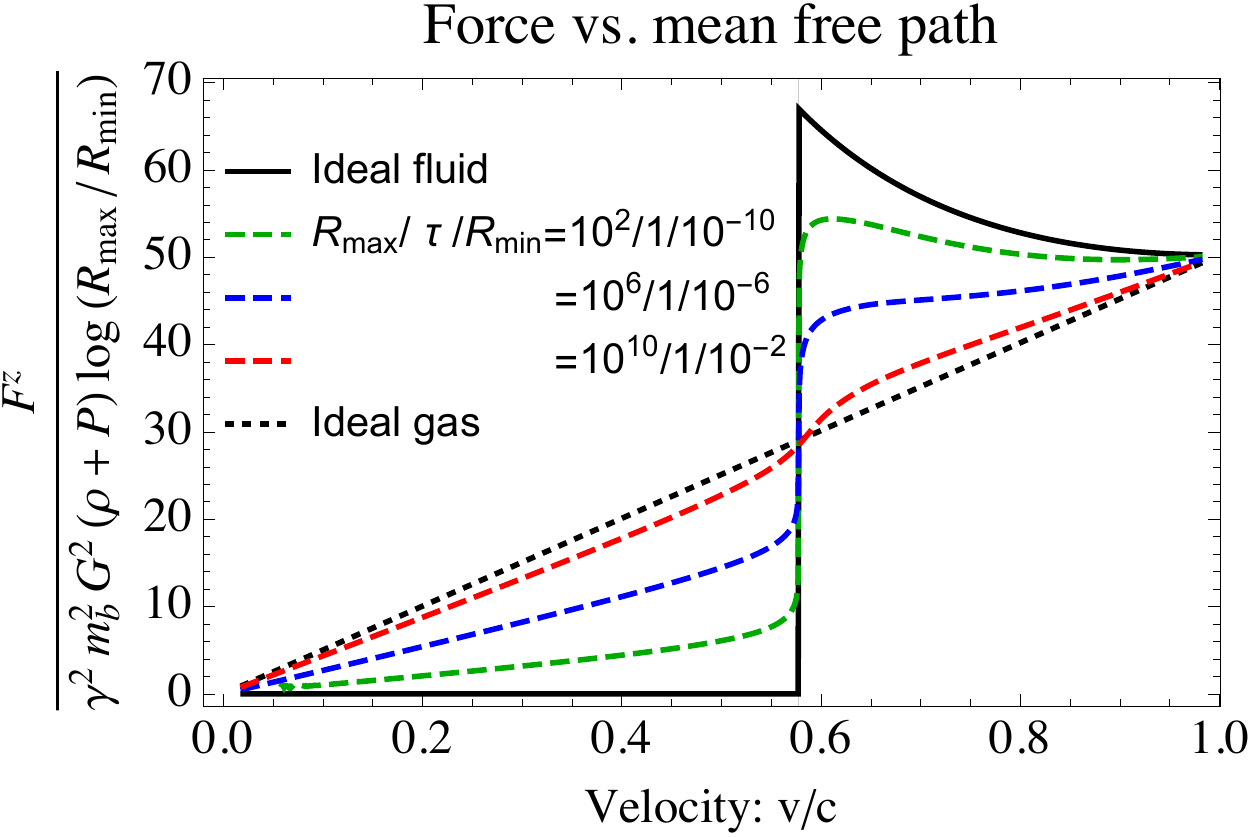}
\includegraphics[width=0.49\textwidth]{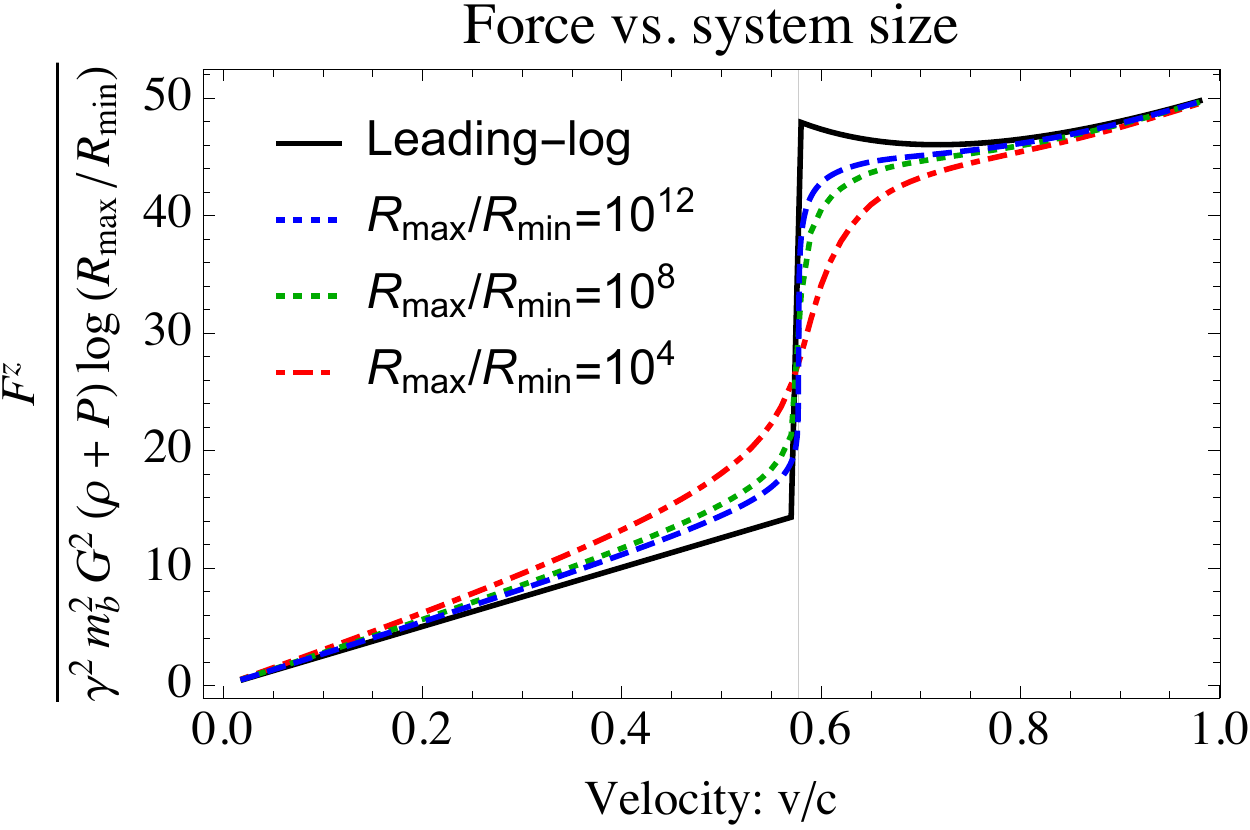}
\caption{
The drag force in interacting kinetic theory. The left panel demonstrates the passage from a medium that is ideal fluid at all scales ($R_{\rm min} \ll \tau $, black solid line) to a medium that is ideal gas at all scales ($R_{\rm max} \ll \tau$, black dotted line). The other lines correspond the systems where the mean free path interpolates between the dimensions of the system $R_{\rm min} < \tau < R_{\rm max}$ keeping the external dimensions of the system fixed $R_{\rm max}/R_{\rm min}=10^{12}$. The right panel demonstrates the effect of the external dimensions of the system. The different lines correspond to different UV- and IR-cutoffs:  
the black solid line corresponds to an infinite system, whereas the other lines correspond to finite systems with different $R_{\rm max}/R_{\rm min}$ ratios with a fixed $R_{\rm max}/\tau = \tau/R_{\rm min}$. 
}
\label{force}
\end{figure}

 Finally, Fig.~\ref{force} shows the full drag force in the interacting kinetic theory, for simplicity, in the massless limit. Because of the Coulomb divergence of the ideal-fluid and ideal-gas regimes the infrared and the ultraviolet regulators need to be implemented to arrive at a finite result.  Here we implement a hard cut-off restricting the $k_\perp$ to an interval $1/R_{\rm max} < k_\perp  < 1/R_{\rm min}$. 
 
 The left panel shows how system with finite mean free path interpolates between
ideal-fluid-like and ideal-gas-like behavior as a function of the mean free
path in a finite system with a constant external dimensions $R_{\rm max}/R_{\rm
	min} = 10^{12}$. The solid black line corresponds to a system that is ideal
fluid at all scales, that is $\tau \ll R_{\rm min}$. The force in ideal fluid
vanishes exactly in the subsonic regime and increases discontinuously at
$v=c_s$. As the interaction rate is reduced the discontinuity is smoothened and
the force is non-zero also for a subsonic projectile. We see that even if the
mean free path is only slightly larger than the size of the projectile, say
$\tau/R_{\rm min} = 10^2$ (green dashed line), there is substantial drag even
in the subsonic regime. Eventually, as the mean free path gets larger and larger
the ideal gas result is approached.

 The right panel demonstrates how the dimensions of the system affect the drag
force. In this figure, the mean free path is chosen to be within the external
dimensions of the system, logarithmically equidistant from the cutoffs $R_{\rm
	max}/\tau = \tau/R_{\rm min}$, and the ratio $R_{\rm max}/R_{\rm min} $ is
varied---for a meaningful comparison the force is normalized by the Coulomb
logarithm $\log R_{\rm max}/R_{\rm min}$. In the limit of infinitely large
system (black line, corresponding to the leading-log expression with $c_X=0$)
the discontinuity arising from ideal fluid dynamics is visible. As the system
is made smaller with reducing $R_{\rm max}/R_{\rm min} $ the discontinuity
rendered continuous by $c_X$. It is noteworthy that even in a system
characterized by a large scale separation of $R_{\rm max}/R_{\rm min}=10^{12}$
the discontinuity is markedly rounded.
 
\section{Comments on Boundary Effects}
\label{sec:boundary}


In the previous sections we assumed that the bullet has traveled in the medium for an infinite time and regulated the long-distance Coulomb divergence by considering a sharp IR-cutoff at $k_\perp = 1/R_{\rm max}$. If the bullet had traveled only a finite time in the medium, that would introduce
a different IR regulator, and lead into different subleading-log contribution at the scale $R_{\rm max}$. Such a scenario was discussed in ideal fluid by Ostriker in~\cite{Ostriker:1998fa}, where it was found that also these effects render the drag a force a continuous function of velocity. 

Without performing a full calculation, we discuss here how such finite boundaries can be 
included in our framework. The situation discussed in \cite{Ostriker:1998fa} is that of a medium that still has an infinite extent but the bullet appears in the unperturbed medium at time $t=0$.

A similar setup can be manufactured in our framework by replacing the energy-momentum tensor of the bullet with a retarded one
\begin{align}
T^{\mu \nu}_{\bh}(\vec x, t) &= \gamma \, m \,\delta(z-v t)\delta^2(\vec x_\perp) v^\mu v^\nu \times \theta(t), \\
\tilde T^{\mu \nu}_{\bh}(\vec k,\omega) &= \gamma \, m \,\frac{-i v^\mu v^\nu}{-\omega + k^z v-i \epsilon}.
\end{align}
Then the (derivative of) the gravitational field of the wake reads
\begin{align}
\partial_\rho h^{wake}_{\mu \nu}(x_\perp =0, z = v t >0 ) &
= 64 i \pi^2 G^2 \gamma m v^\alpha v^\beta \int \frac{d^4 k}{(2\pi)^4} e^{-i t (\omega - i k^z v )} k_\rho  G^{dressed}_{\mu \nu,\alpha \beta}(k)\frac{-i v^\mu v^\nu}{-\omega + k^z v-i \epsilon},
\end{align}
and the force 
\begin{align}
F^z=\frac{m^2\gamma^2}{2} \int \frac{\text{d}^2 k_\perp}{(2\pi)^2} \frac{\text{d}k^z}{2\pi}k^z & \frac{G^{\dressed} }{-\omega + k^z v-i \epsilon}
+m^2\gamma^2\int \frac{\text{d}^2 k_\perp}{(2\pi)^2} \frac{\text{d}k^z}{2\pi}\left( \omega - k^z v\right) \frac{\tilde G^{\dressed} }{-\omega + k^z v-i \epsilon}
\label{Eq:Fz}
\end{align}
with 
\begin{align}
\tilde G^{\dressed} = G^{\dressed}_{z\mu \alpha\beta}\tilde v^{\mu }v^{\alpha}v^{\beta}, \quad \quad \tilde v^{\mu} = \{ 1,0,0,-v \}
\end{align}
We leave further development and the evaluation of this expression for specific media to future studies.

\section{Discussion and Outlook }
\label{sec:conclusions}


In this paper we have calculated the drag  force in interacting relativistic medium. 

Unlike in the previous 
works on this subject, we did not restrict ourselves to any particular kinematic regime, but presented a 
generic calculation that is valid at all masses and interaction rates. We showed that our techniques easily allow us to recover the previously known 
limits (relativistic and non-relativistic free-streaming gas, ideal fluid) as well as the interpolations between them. 
The results for viscous fluid, relativistic free-streaming gas and for interacting gas with a finite mean free path 
were presented 
for the first time.
We studied interacting kinetic theory in a simple relaxation-time approximation and found that 
the important qualitative and quantitative features of the dynamical friction going beyond the leading-log approximation are reliably captured by the viscous fluid dynamics. Therefore, we believe that these features are unlikely to change if 
we consider other, more realistic models of particle interactions.

The calculation that we have presented here is merely the first attempt to tackle the 
dynamical friction problem with these new techniques. There are several obvious 
ways how one can improve these calculations, make them more precise, and even better suited 
for the practical problems in astrophysics. First, in this paper we have merely studied the 
Newtonian approximation. While it is sufficient for many practical problems, knowing 
the post-Newtonian corrections would be desired. While they have been considered
in specific limits, a generic understanding of them in the case of generic masses and 
interaction rates is missing. One can, in fact, access the post-Newtonian corrections 
with our techniques by expanding to the next-to-leading order in the graviton loops, 
treating the gravity as an effective field theory~\cite{Donoghue:2017pgk}. Another useful application 
would be to understand the boundary effects on the dynamical friction appearing at scale $R_{\rm max}$. While we made the 
first step in Sec~\ref{sec:boundary}, there is still lots of room for progress, where one can 
consider the spatial-temporal boundaries~\cite{Ostriker:1998fa} 
(invoking, \emph{e.g.}, mirror charges). Also, more exotic media can be discussed (for recent work see, \emph{e.g.} \cite{Berezhiani:2019pzd}). Finally, one can also study dynamical friction
in non-linear movement, using the same formalism but changing appropriately 
the energy momentum tensor (for similar works using classical techniques 
see~\cite{Kim:2007zb}). 

Our results can be applied to a large number of astrophysical systems such as
galactic dynamics with interacting dark matter (see \emph{e.g.}~\cite{Kahlhoefer:2015vua})
and the interaction of
primordial black holes with neutron stars and white dwarfs. 
In particular, in setting limits on primordial black holes as a dark-matter component,
the capture rate by compact objects is determined by dynamical friction in interacting
and relativistic setting. Taking this into account, it will interesting to revise the previous 
bounds claimed by~\cite{Capela:2012jz,Graham:2015apa}.
The effects of dynamical friction in extreme conditions
are also in an important role in studies of the hypothetical stable 
TeV-scale black holes~\cite{Giddings:2008gr}. 
With these 
new techniques and results here presented we expect to tackle these and 
many other questions. 

\vspace{1.0cm}
\noindent
{\bf Acknowledgments.} We are grateful to Sergey Sibiryakov for numerous 
discussions and insights and collaboration at the early stages 
of this project as well as the comments on the manuscript. We are also grateful to Anton Rebhan for useful 
comments on the manuscript and Urs Wiedemann for useful discussions. A. Soloviev was supported by the Austrian Science Fund 
(FWF) doctoral program W1252.
\appendix

\section{Graviton propagator}
\label{App:graviton}


For a generic linear perturbation of the energy momentum tensor $\delta T^{\mu\nu}$, the linearized Einstein equations in harmonic gauge ($\partial_\mu g^{\mu} _{\phantom{\mu}\nu}-\frac{1}{2}\partial_\nu g^{\mu}_\mu=0$) take the following form \cite{Donoghue:2017pgk}%
\footnote{Note that our notation differs from that of \cite{Donoghue:2017pgk} by $h_{here} = 8\pi G h$.

In addition \cite{Donoghue:2017pgk} uses mostly minus metric such that $\Box_{here}=-\Box$.}
\begin{equation}
\Box \bar{h}_{\mu\nu}\equiv\Box (h_{\mu\nu}-\frac{1}{2}\eta_{\mu\nu}h)=\frac{1}{2}(\eta_{\mu\alpha}\eta_{\nu\beta}+\eta_{\mu\beta}\eta_{\nu\alpha}-\eta_{\alpha\beta}\eta_{\mu\nu})\Box h^{\alpha\beta}=16 \pi G  \delta T_{\mu\nu}. 
\label{eq:linE}
\end{equation}

We define a Green function for the above equation through,
\begin{align}
\frac{1}{2}(\eta_{\mu\alpha}\eta_{\nu\beta}+\eta_{\mu\beta}\eta_{\nu\alpha}-\eta_{\alpha\beta}\eta_{\mu\nu}) \Box G_{\grav}^{\alpha\beta,\gamma\delta}(x,x') = 8\pi G(\eta_\mu^\gamma \eta_\nu^\delta+\eta_\mu^\delta \eta_\nu^\gamma )\delta^{(4)}(x-x')
\end{align}
which is conveniently solved in Fourier space
\begin{equation}
\label{eq:gravition}
G_{\grav}^{\mu\nu,\alpha\beta}(x,x') = 8\pi G \frac{(\eta^{\mu\alpha}\eta^{\nu\beta}+\eta^{\mu\beta}\eta^{\nu\alpha}-\eta^{\alpha\beta}\eta^{\mu\nu})}{-(\omega+i \epsilon)^2 + k^2} \equiv 8\pi G \frac{I^{\mu \nu, \alpha \beta}}{-(\omega+i \epsilon)^2 + k^2},
\end{equation}
such that 
\begin{equation}
h_{\mu \nu}(x) =  \int d^4 x' G_{\mu\nu,\alpha\beta}^{\grav}(x-x') T^{\alpha\beta}.
\end{equation}

\section{Numerator structures }
\label{RS}
For completeness, we display here the full numerator structures appearing in Eq.~\ref{Gdressedfreestream}
\begin{align}
R=&-2 p \sqrt{k_\perp^2+k_z^2}  \Big{[}2 k_\perp^8  (p^4  (7 v^4-10 v^2+15 )-15 (p^0)^4  (v^2+1 )^2 )\nonumber\\
&+k_\perp^6 k_z^2  (8 p^4  (7 v^4-10 v^2+15 )+15 p^2   (p^0)^2 v^2  (v^4+20 v^2+16 )+60 (p^0)^4  (v^6+7 v^4-2 ) )\nonumber\\
   &+k_\perp^4 k_z^4  \Big{(}12 p^4  (7 v^4-10 v^2+15 )+5 p^2 (p^0)^2 v^2  (-33 v^4+80
   v^2+144 )\nonumber\\
   &-45 (p^0)^4  (v^8+16 v^6-12 v^4-8 v^2+4 ) \Big{)}\nonumber\\
   &+4 k_\perp^2 k_z^6  \Big{(}2 p^4  (7 v^4-10 v^2+15 )-5 p^2 (p^0)^2 v^2  (5 v^4+5 v^2-36 )\nonumber\\
   &+15 (p^0)^4
    (6 v^8-7 v^6-5 v^4+8 v^2-2 ) \Big{)}\\
    &+2 k_z^8  (p^4  (7 v^4-10 v^2+15 )+20 p^2 (p^0)^2 v^2  (2 v^4-5 v^2+6 )-15 (p^0)^4  (2 v^4-3 v^2+1 )^2 ) \Big{]}\nonumber\\
S&= -15 k_z p^0 v  \Big{[}k_\perp^8  (p^4  (v^4+2 )+4 p^2 (p^0)^2  (5 v^2+1 )+2 (p^0)^4  (v^2+1 )^2 )\nonumber\\
&+2 k_\perp^6 k_z^2  (p^4  (v^4-2
   v^2+4 )-p^2 (p^0)^2  (v^6+20 v^4-22 v^2-8 )-2 (p^0)^4  (v^6+7 v^4-2 ) )\nonumber\\
   &+3 k_\perp^4 k_z^4  (p^4  (v^2-2 )^2+2 p^2 (p^0)^2  (3 v^6-10 v^4+2
   v^2+4 )+(p^0)^4  (v^8+16 v^6-12 v^4-8 v^2+4 ) )\nonumber\\
   &+4 k_\perp^2 k_z^6  (v^2-1 )  (p^4  (v^2-2 )+p^2 (p^0)^2  (3 v^4+3 v^2-4 )+(p^0)^4
    (-6 v^6+v^4+6 v^2-2 ) )\nonumber\\
    &+2 k_z^8  (v^2-1 )^2  (p^2+(p^0)^2  (1-2 v^2 ) )^2 \Big{]}
\end{align}
and 
\begin{align} \nonumber
f_{\rm sound}(\omega, \vec k) 
&= f_{\rm sound}^{\rm ideal} +
\frac{\rho_0+P_0}{2 }\Big{\{}
4 i \omega \Big{[}k^2 \left(c_s^2 \left(v^4+3\right)+2 \left(v^2+1\right)\right)-2   k_z^2 v^2 \left(c_s^2 \left(v^2+3\right)+v^2+1\right)
\\&+\frac{4 k_z^2 v^3 \left(c_s^2 k_z^2 v+\omega  (2 k_z+v \omega
	)\right)}{k^2}-\frac{4 k_z^4 v^4   \omega ^2}{k^4}-8 k_z v \omega -\left(v^4+2 v^2-3\right) \omega ^2\Big{]}\eta_s \nonumber \\
&+i  \omega\left(v^2-3\right)   \big{[}k^2 \left(v^2+1\right)-\omega  \left(8
k_z v+\left(v^2-3\right) \omega \right)\big{]}\Gamma_s \Big\} 
\\
f_{\rm shear}( \omega,\vec k) &=-\frac{8(\rho_0+P_0)}{k^4}  v^2 \omega  k_\perp^2  \left( k^2+ i\eta _s \left(2 k^2 k_z v-k_z^2 v^2 \omega \right)\right)
\end{align}
\section{Fluid-dynamic propagator }
\label{App:fluid}
\label{fluid-dynamics}
In this Appendix we consider the fluid dynamical response function studied extensively in 
literature~(see \emph{e.g.}, \cite{Baier:2007ix}). The starting point for the (1st order) relativistic dissipative fluid dynamics is the conservation of energy and momentum
\begin{align}
\nabla_\alpha T^{\alpha \mu} = 0,
\label{Eq:econs}
\end{align}
combined with a gradient expansion of the energy-momentum tensor including the terms containing at most one derivative of the flow and density fields $u^\mu(x)$ and $\rho(x)$. The most generic $T^{\mu\nu}$ is given by the constitutive equation
\begin{align}
T^{\mu\nu}&=(\rho + P)u^\mu u^\nu+ P g^{\mu\nu}-2\eta \sigma^{\mu\nu}-\zeta\nabla_\alpha u^\alpha\Delta^{\mu\nu},\nonumber\\
\sigma^{\alpha\beta}&=\frac{1}{2}\Delta^{\mu\alpha}\Delta^{\nu\beta}\Big{(}\nabla_\mu u_\nu+\nabla_\nu u_\mu -\frac{2}{3}\nabla_\alpha u^\alpha g_{\mu\nu}\Big{)}\equiv \langle\nabla^\alpha u^\beta \rangle,
\label{Eq:const}
\end{align} 
where $\eta$ and $\zeta$ are the shear and bulk viscous transport coefficients, and
where $\Delta^{\mu\nu}\equiv u^\mu u^\nu+g^{\mu\nu}$ is a spatial projector. We work in the Landau frame, \emph{i.e.}, $u^\mu \sigma_{\mu\nu}=0$. 

In the following we will compute the linear response to a static and homogenous fluid with $\rho = \rho_0$ and $u_0^{\mu} =(1,0,0,0)$ caused by a linear perturbation of the metric $ g_{\mu\nu}=\eta_{\mu\nu}+ h_{\mu\nu}$.
The density perturbation is parameterized  via $\rho = \rho_0+\delta \rho(x)$ and the velocity perturbation as $u^{\mu}(x^\mu) = u^{\mu} + v^\mu (x^\mu)$, with $v^0 =0$.

We now compute the linear response of the energy density and velocity induced by a plane-wave gravitational-field perturbation directed along the $\hat l$-direction
\begin{equation}
g^{\mu \nu}(x^\mu) = \eta^{\mu\nu} + h^{\mu \nu} e^{-i\omega t +i \vec k  \cdot \hat l}.\end{equation}
Inserting the constitutive equation \ref{Eq:econs} to energy conservation equation \ref{Eq:econs} and using the above ansatz, gives four equations that can be solved for the perturbations
\begin{align}
\delta \rho&=-\frac{1}{2} \frac{  (\rho_0+P_0)( k^2 h_{00}+2
	\omega k h_{0l}  +\omega^2 h_{ij}\delta^{ij})+2 i \eta \omega k^2  (h_{mm} +h_{nn})
}{ \omega ^2- c_s^2 k^2+ i \Gamma_s \omega k^2  },\nonumber\\   
v_m&=
\frac{\eta\omega  k h_{ml} +i  (\rho_0+P_0)\omega h_{0m}}{\eta  k^2-i \omega
	(\rho_0+P_0)},\nonumber\\
v_n&= \frac{  \eta \omega k h_{nl}+i 
	(\rho_0+P_0)\omega h_{0n}}{\eta  k^2-i \omega  (\rho_0+P_0)},\nonumber\\
v_l&=
- \frac{1}{2}\frac{(c_s^2 \omega k 
	h_{ij}\delta^{ij}+\omega k h_{00} +2 \omega^2 h_{0l} -i \Gamma_s \omega^2  k  h_{ij}\delta^{ij} )
	+2i \eta \omega^2 k (h_{mm}+h_{nn})
}{ \omega^2 -c_s^2 k^2+i
	\Gamma_s \omega k^2}
\end{align}
where $\Gamma_s k^2$ is the sound attenuation length with $\Gamma_s =(\frac{4}{3}\eta+\zeta)/(\rho_0+P_0)$.

Defining the fluid dynamical response function by%
\footnote{Note that in the calculation we have chosen not to treat independently the off-diagonal elements of $h_{\mu\nu}$. However in the calculation of the force the two components get counted separately and for that reason we must divide the off-diagonal terms by a factor of two to avoid double counting. }
\begin{align}
\delta T^{\mu \nu}(x) &= \int d^4 x' G_{\medium}^{\mu\nu,\alpha \beta}(x,x') h_{\alpha\beta}(x') \\
\frac{\delta T^{\mu \nu}(\omega, \vec k)}{\delta h_{\alpha\beta}} &=G_{\medium}^{\mu\nu,\alpha \beta}(\omega,\vec k) \times
\left\{ \begin{array}{c} 
1, \quad \alpha = \beta \\
1/2, \quad \alpha \neq \beta
\end{array}
\right.
\end{align}
one can straightforwardly find the components of the correlation function
\begin{align}
\label{Ghydro}
G_{\medium}^{00,00} &= -\frac{(P_0+\rho_0)}{2}\frac{k^2 }{- (\omega +i \Gamma_s k^2)^2+c_s^2 k^2}-\rho_0  \\
G_{\medium}^{0x,0x} &=-\frac{(P_0+\rho_0)}{2} \frac{i \eta\omega}{i\omega-\frac{\eta k^2}{\rho_0+P_0}}-\frac{P_0}{2}  \label{shearmode}\\
G_{\medium}^{xy,xy} &= \frac{i \eta \omega }{2}-\frac{P_0}{2} \label{tensormode}
\end{align}
The first one corresponds to the sound channel, the second the shear channel and the last the tensor channel. The rest of the components can be obtained by the repeated use of the Ward identities, $\nabla_\mu T^{\mu\nu}=0$, that relate components of the Green function within the different channels.

\section{Extraction of Transport Coefficients from Kinetic Theory}
\label{App:transport}
\label{hydro-coeff}
\begin{figure}
	\centering
	\includegraphics[width=.49\textwidth]{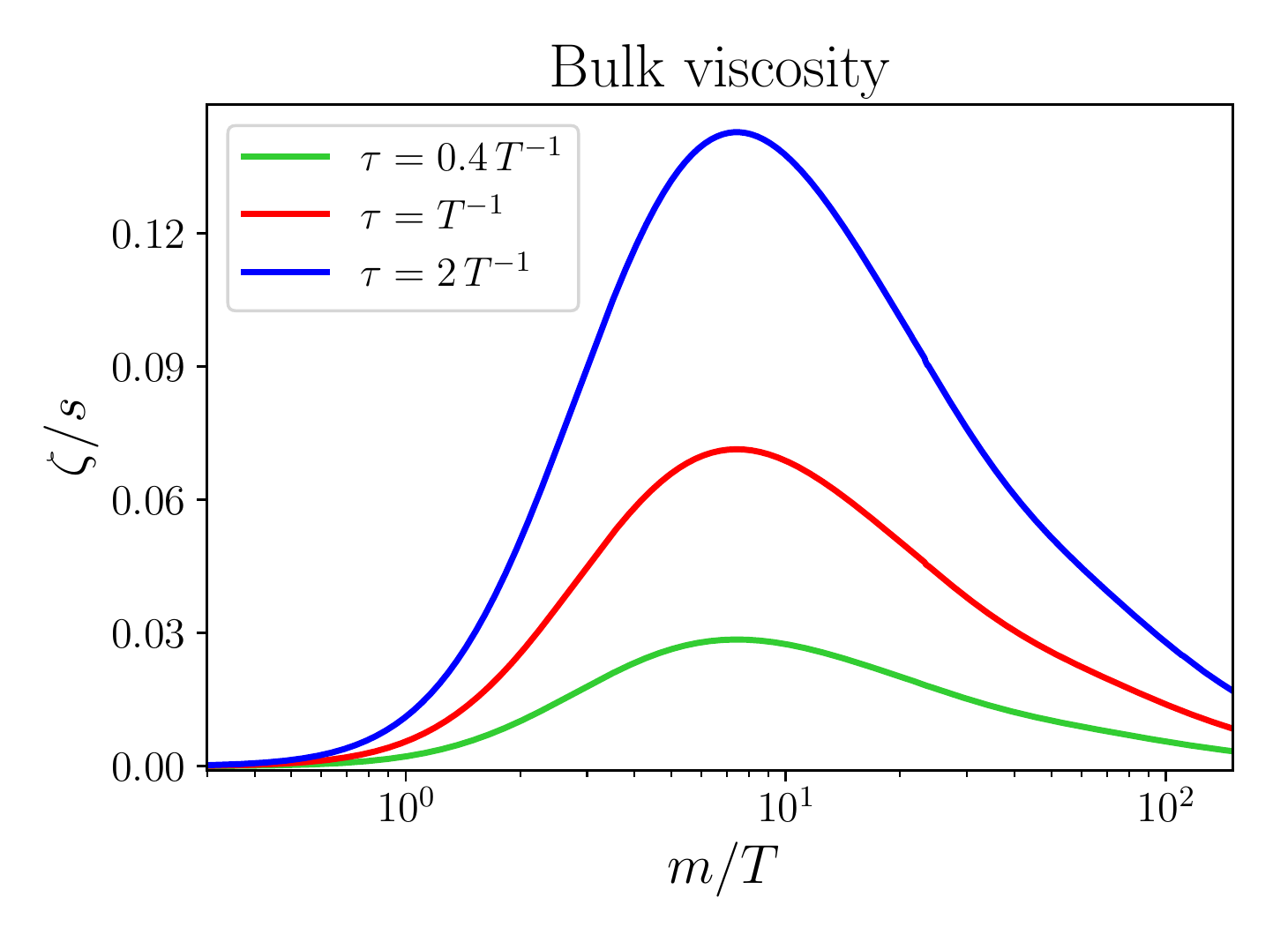}
		\includegraphics[width=.49\textwidth]{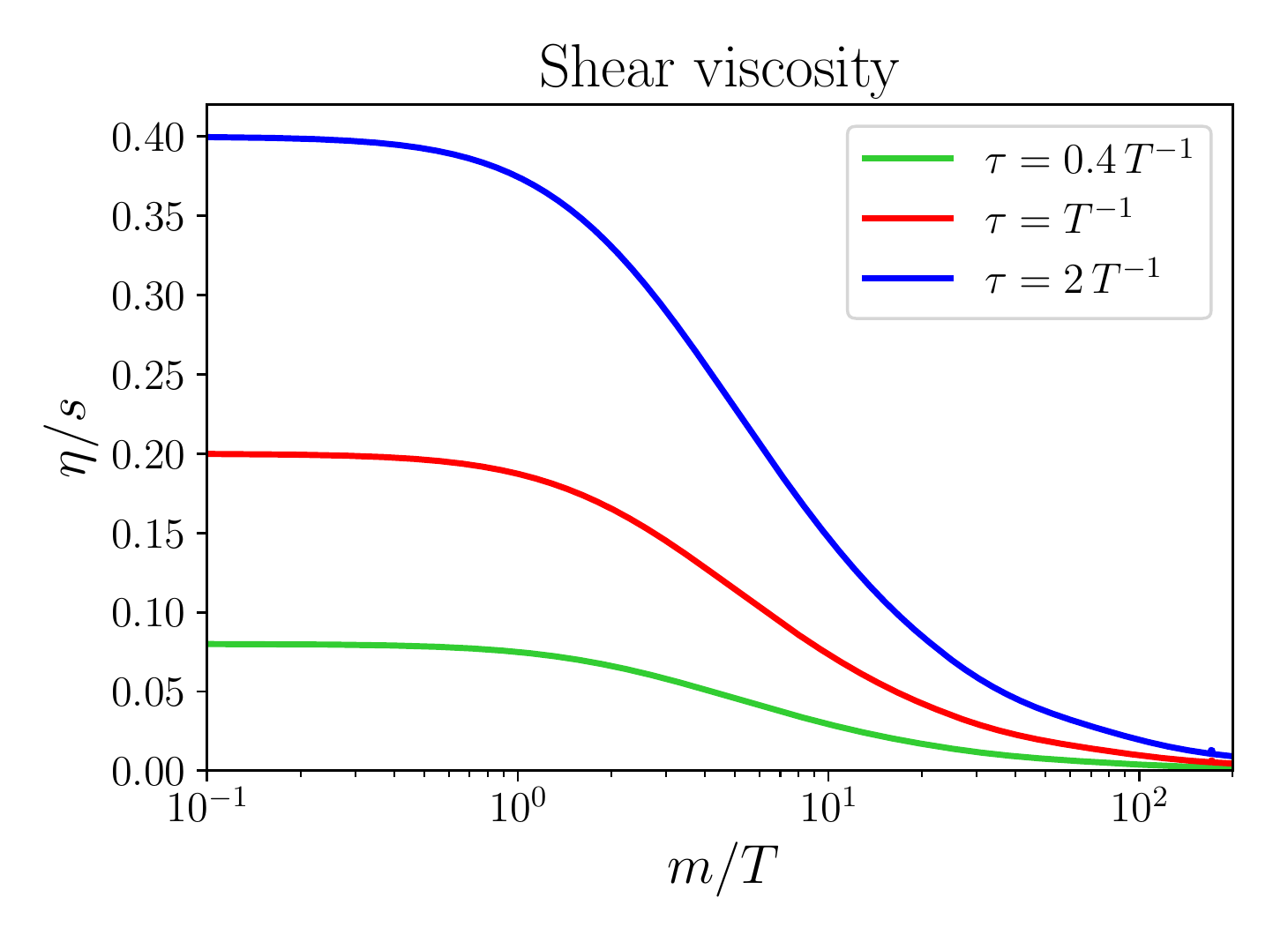}
	\caption{Matching of the bulk viscosity onto the parameters of the full kinetic
	theory. The bulk viscosity, normalized to the entropy density is maximized
	around $m \sim 10T$. In the limit of zero mass of the gas  particle both
	viscosities recover a well known conformal limit, $\zeta \to 0$ and $\eta \to
	s T \tau/5$.}
	\label{fig:transport}
\end{figure}
In this appendix we match the the transport coefficient of the effective theory
of viscous liquid onto the variable of full kinetic theory, namely the
relaxation time $\tau$ and the mass of the gas particle $m$ (for a very similar 
calculation see~\cite{Czajka:2017wdo}).  
The result for the
the Maxwell Boltzmann distribution reads

\beq
\eta & = & \int dp \frac{e^{-\frac{\sqrt{m^2 + \vec p^2} }{T}} p^6 \tau}{30 \pi^2 (m^2 + \vec p^2)} \\ \label{eq:zeta}
\zeta & = & \int \frac{dp e^{-\frac{\sqrt{p^2 + m^2}}{T}} m p^4 \tau (p^2 K_2(m/T) - 3 m TK_3(m/T)}{9 (m^2 + p^2)
\pi^2 (mK_2(m/T) + 3 T K_3(m/T)}
\eeq 
where we have used to the specific heat 
\beq
C_V= \frac{d\rho}{T d \log T}  = \frac{m^3 (\frac{m}{T} K_2(m/T) + 3K_3(m/T)}{2 \pi^2}
\eeq 

We plot both these transport coefficients on Fig.~\ref{fig:transport}. Note that in the conformal limit 
we recover the well-known expression for $\eta/(\rho + P) \to \tau /5$, while the bulk viscosity 
vanishes. 



\bibliography{dynfricbib}
\bibliographystyle{JHEP}

\end{document}